\documentclass[twocolumn,
nofootinbib,superscriptaddress,aps,10pt,longbibliography]{revtex4-2}

\usepackage{bbold}

\usepackage[usenames,dvipsnames]{xcolor}
\usepackage{amsmath}
\usepackage{amsthm, amssymb}
\usepackage{enumitem}
\usepackage{slashed}
\usepackage{graphicx}
\usepackage{tikz}
\usetikzlibrary{calc,fadings,decorations.pathreplacing,shapes,shapes.multipart,arrows,shapes.misc,intersections,positioning,patterns}
\usepackage{bm}
\usepackage{dsfont}
\usepackage{changepage}
\usepackage[normalem]{ulem}
\usepackage{array}

\definecolor{bluepurple2}{rgb}{0.06,0,0.6}
\usepackage[colorlinks=true,citecolor=blue,linkcolor=bluepurple2]{hyperref}

\usepackage{bm}
\renewcommand{\vec}[1]{\boldsymbol{\mathbf{#1}}}

\graphicspath{{./}{./images/}}

\newcommand{\bit}{\begin{itemize}}
\newcommand{\eit}{\end{itemize}}

\newcommand{\f}{\frac}
\renewcommand{\>}{\right\rangle}
\newcommand{\<}{\left\langle}
\newcommand{\ba}{\begin{align}}
\newcommand{\ea}{\end{align}}
\newcommand{\be}{\begin{equation}}
\newcommand{\ee}{\end{equation}}
\newcommand{\bi}{\begin{itemize}}
\newcommand{\ei}{\end{itemize}}
\newcommand{\lf}{\left(}
\newcommand{\ri}{\right)}
\newcommand{\dd}{\mathrm{d}}

\newcommand{\ZZ}{\mathbb{Z}_2}
\newcommand{\Apat}{A_\text{patched}}

\begin{document}
\date{\today}

\newcommand{\bra}[1]{\< #1 \right|}
\newcommand{\ket}[1]{\left| #1 \>}

\title{Worldsheet patching, 1-form symmetries, and ``Landau-star'' phase transitions}

\author{Pablo Serna}
\affiliation{Current address: Genaios, Valencia, Spain}
\affiliation{Departamento de F\'isica -- CIOyN, Universidad de Murcia, Murcia 30.071, Spain}
\author{Andr\'es M. Somoza}
\affiliation{Departamento de F\'isica -- CIOyN, Universidad de Murcia, Murcia 30.071, Spain}
\author{Adam Nahum}
\affiliation{Laboratoire de Physique de l'\'Ecole Normale Sup\'erieure, CNRS, ENS \& Universit\'e PSL, Sorbonne Universit\'e, Universit\'e de Paris, 75005 Paris, France.}

\date{\today}

\begin{abstract}
The analysis of phase transitions of gauge theories  has relied heavily on
simplifications that arise at the boundaries of phase diagrams,  
where  certain  excitations are forbidden.  
Taking 2+1 dimensional $\mathbb{Z}_2$ gauge theory as an example, the simplification can be visualized geometrically: 
on the phase diagram boundaries 
the partition function is an ensemble of \textit{closed membranes}.
More generally, however, the membranes have ``holes'' in them, representing  worldlines of virtual anyon excitations.
If the holes  are of a finite size, then typically they do not  affect the universality class, but 
they destroy  microscopic (higher-form) symmetries and microscopic (string) observables.
We demonstrate how these symmetries and  observables can be restored using a ``membrane patching''  procedure, which maps the ensemble of membranes back to an ensemble of closed membranes. (This is closely related to the idea of gauge fixing in the ``minimal gauge'', though not equivalent.)
Membrane patching makes the emergence of higher symmetry concrete.
Performing patching in a Monte Carlo simulation with an appropriate algorithm,
we show that it gives access to numerically useful observables.  For example, the confinement transition can be analyzed using a correlation function that is a power law at the critical point.
We analyze the quasi-locality of the patching procedure 
and  discuss what happens at a self-dual multicritical point in the gauge-Higgs model, 
where the lengthscale $\ell$ characterizing the  holes diverges.
The patching approach described here generalizes to many other statistical ensembles with a  representation  in terms of fluctuating membranes or loops, in various dimensions, and related constructions could be implemented in experiments on quantum devices.
\end{abstract}
 
 \maketitle

\section{Introduction}
\label{sec:intro}

It is a remarkable fact that many topological phase transitions, which  lack any local order parameter, can be described using Landau-Ginsburg theory anyway. 
The Landau theory is formulated in terms of a ``fictitious'' order parameter, 
which does not exist as a local observable of the original theory. 
Canonical examples include Higgs transitions in discrete gauge theories,
and in three dimensions also  their confinement transitions \cite{wegner1971duality,fradkin1979phase}. 
The simplest case is the 3D gauge-Higgs model with $\ZZ$ gauge field and $\ZZ$ matter, where both the Higgs and confinement transitions are ``Ising$^*$''  transitions \cite{lammert1993topology,senthil2002microscopic,grover2010quantum,misguich2008quantum,moessner2001ising,slagle2014quantum}.
These are not Ising transitions, but they can be largely understood in terms of the latter.
In the deep infra-red, the Ising$^*$ fixed point, or more generally a ``Landau$^*$'' fixed point, is related to the standard Landau fixed point by  orbifolding: i.e. by a gauging procedure that eliminates non-gauge-invariant operators, including the Landau order parameter, from the spectrum. 
(We will review ways of thinking about Landau$^*$ transitions below.)

A crucial point about the relationship between the gauge theory and the Landau theory is that in general it is emergent, rather than microscopic.
In other words, the relation is meaningful only after coarse-graining beyond some characteristic lengthscale $\ell$.
For a Higgs transition, $\ell$ separates a regime at shorter scales, where gauge fluctuations are nontrivial, 
from a regime at larger scales where (heuristically) the gauge field can be treated as flat --- i.e. locally equivalent to pure gauge. Once we have an effective model where nontrivial gauge fluctuations have been eliminated, the relation to a Landau-like theory is direct:  we can locally choose a gauge where the gauge field is trivial, so that the interactions of the Higgs fields resemble those in a Landau theory \cite{wegner1971duality, fradkin1979phase}.  The emergence of the Landau$^*$ description is also tied to the emergence of a ``higher'' symmetry \cite{nussinov2009symmetry,kapustin2017higher,gaiotto2015generalized,wen2019emergent,hastings2005quasiadiabatic}, associated with emergent string operators, as discussed below.

Anyon excitations of the deconfined phase give an alternative perspective.
For example, the Higgs transition of 2+1D $\ZZ$ gauge theory is the condensation of the ``{\it e}'' anyon.
The lengthscale $\ell$ is related to the mass scale for \textit{other} anyons, which braid nontrivially with {\it e}, but which can be neglected in the infra-red.
At the longest scales,  the dynamics of the {\it e} particle then resemble those of a local boson in a Landau theory.

In the case where $\ell$ is suppressed all the way to zero, 
the ``duality'' 
between the transition in the gauge theory and the Landau transition 
holds at the microscopic level
\cite{wegner1971duality}.
(This may be extended to small $\ell$ within perturbation theory \cite{wegner1971duality, fradkin1979phase}.)
However, the scale $\ell$ may be arbitrarily large;
for example we can approach a  phase transition where $\ell$ diverges (and where the Landau$^*$ description breaks down).
In general the Landau$^*$ description and the associated higher symmetry emerge only after a nontrivial renormalization group flow.

The renormalization group pictures above are heuristic.
Ref.~\cite{somoza2021self} outlined how to make the emergence precise and ``constructive'' using spacetime configurations.
This method defines the emergent observables sufficiently concretely that they can be measured in a simulation and their locality properties can be analyzed theoretically. 
Here we develop this approach in full
and implement it in a Monte Carlo simulation. 
Our aims are twofold:

First, to develop a geometrical understanding of
how the Landau$^*$ description
--- with its ``fictitious'' order parameter --- emerges as we go from the microscopic scale to scales $\gg \ell$.
In a more formal language, this is also  the emergence 
\cite{wen2019emergent,somoza2021self,iqbal2022mean,
mcgreevy2023generalized,
cherman2023emergent,
pace2023exact}
of a one-form symmetry  \cite{nussinov2009symmetry,kapustin2017higher,gaiotto2015generalized} --- a symmetry group generated by string operators. (Here, these string operators are associated with anyons \cite{kitaev2003fault,hastings2005quasiadiabatic,wen2019emergent}.) 
We start from the 
geometrical representation of the partition function of a lattice gauge theory as an ensemble of \textit{membranes}  \cite{huse1991sponge,gujrati1989critical,gregor2011diagnosing,kogut1979introduction,somoza2021self}
and use a ``patching'' procedure to construct the emergent observables.
The general idea was outlined in Ref.~\cite{somoza2021self} and is developed here.
The procedure is not limited to lattice gauge theory, and could be applied to many other statistical mechanics ensembles formulated in terms of strings or membranes (we will discuss some examples).

Second, we aim to show that the approach based on ``patching'' membranes is \textit{useful}, in Monte Carlo simulations, for analyzing topological phase transitions. We argue that constructing the emergent ``dual'' order parameter and its correlators gives a  more efficient way of studying the confinement transition than looking at standard local correlators.
It is worth noting that some standard diagnostics for the transition, such as the Fredenhagen-Marcu order parameter \cite{fredenhagen1986confinement,gregor2011diagnosing}, have an exponentially bad signal-to-noise problem, which is not the case here.

The patching-based approach outlined below 
is related to an idea of Fradkin and Shenker \cite{fradkin1979phase}, 
who argued that a two-point correlator for Higgs fields can be defined, within the perturbative regime ($\ell\ll 1$), 
by gauge fixing in the ``minimal gauge''. 
Gauge fixing can be formulated as a geometrical problem 
similar to  the membrane patching proposed here but with some important differences (concerning both  the nature of the patching algorithm and the way excitations on the scale of the system size are treated).
The present ``geometrical'' approach makes it possible to go beyond the perturbative regime of small $\ell$, 
and to construct emergent string operators 
(which in general are not accessible simply by gauge fixing). 
We use concepts from geometrical critical phenomena to analyze the quasilocality of our patching procedure, showing e.g. that it can be used to diagnose the deconfined phase and to study the relevant phase transition out of this phase even if $\ell$ is much larger than the lattice spacing.

\begin{figure}
    \centering
    \includegraphics[width=1.0\linewidth]{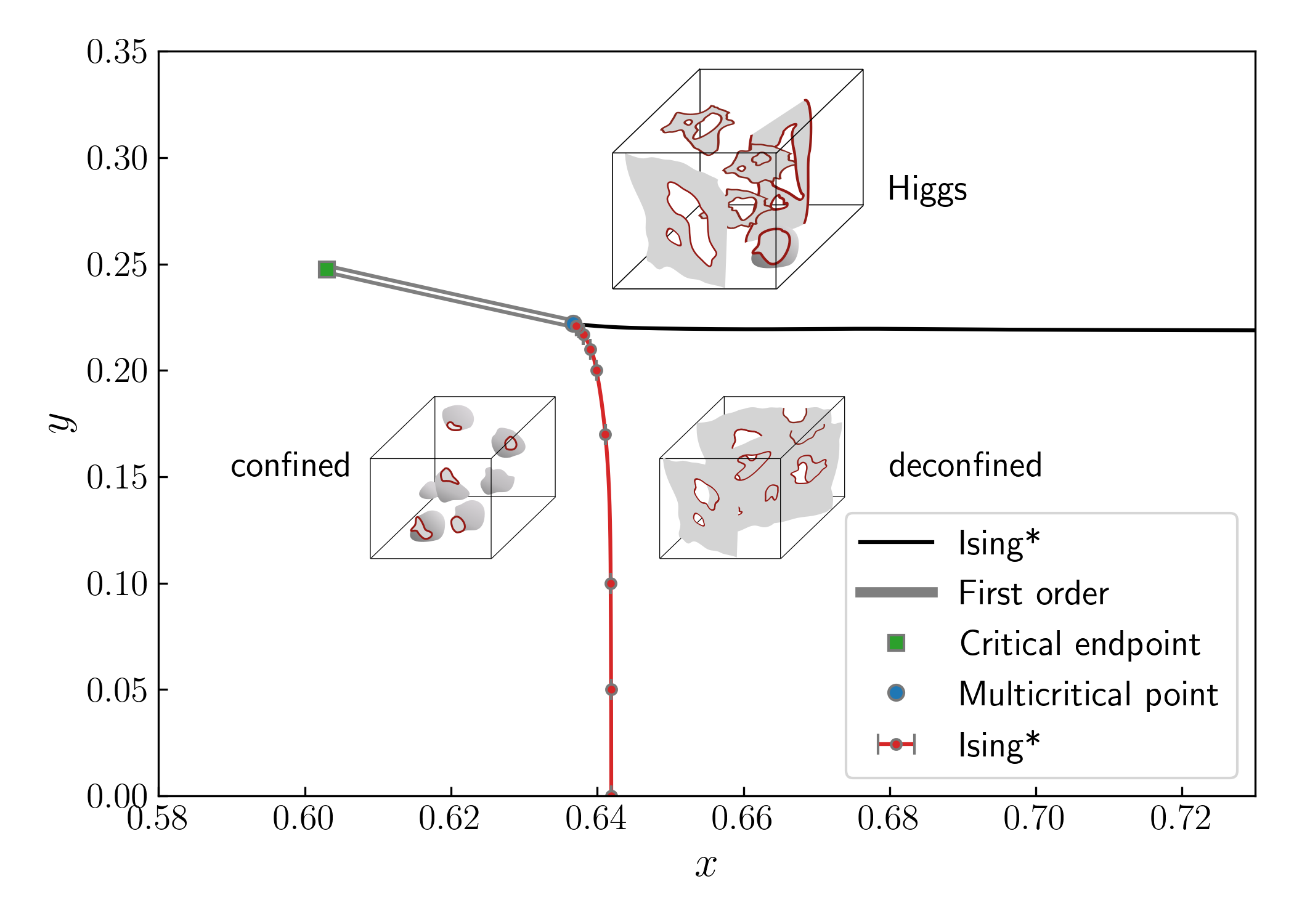}
    \caption{Phase diagram of the $\ZZ$ gauge--Higgs model in 3D, showing the confinement transition line (red), the Higgs transition line (black), the self-dual multicritical point (MCP) where they meet, and a segment of first order coexistence extending from the MCP (double line).  The model's  partition function is shown in Eq.~\ref{eq:Z3Dgaugerep}, with $x,y$ defined in Eq.~\ref{eq:definexy}. 
    The partition function is expressed in terms of classical membranes in  Eq.~\ref{eq:partitionfunctionmembranes}, where  $-\log x$ is the bare membrane tension and $-\log y$ is the bare line tension of membrane boundaries.}
    \label{Fig:phasediagram}
\end{figure}

The patching approach also sheds light on topological phase transitions that are \textit{not} Landau$^*$-like and which are far less understood, including the self-dual point where the two Ising$^*$ lines in Fig.~\ref{Fig:phasediagram} meet  \cite{vidal2009low,TupitsynTopological,somoza2021self}, which is itself a scale-invariant multicritical point~\cite{somoza2021self,bonati2022multicritical,oppenheim2023machine}. We discuss the fate of the string operators at this multicritical point (MCP). 
Interestingly, although they become nonlocal, there is a sense in which this nonlocality is weak.
We discuss why this MCP is challenging to analyze using continuum field theory
(despite the existence of a continuum Chern-Simons formulation for $\mathbb{Z}_2$ topological order).

 We end this introduction with a heuristic overview of the geometrical picture. 
$\ZZ$ gauge theory can be formulated in terms of unoriented membranes
\cite{huse1991sponge,gujrati1989critical,gregor2011diagnosing}.
(In the quantum interpretation, these membranes represent  worldsheets traced out by  flux lines in the 2+1D theory, as they evolve in imaginary time: see e.g.  Fig.~6 of \cite{somoza2021self}).
For us, the key distinction is between membranes/worldsheets that are \textit{closed} and those that have ``holes'' in them  (to be made more precise below): see Fig.~\ref{Fig:phasediagram}.

When all membranes are closed, the mapping between the gauge theory and a Landau theory (here Ising) is almost trivial.
The membranes are simply reinterpreted as domain walls in an Ising order parameter.\footnote{For the moment we work in the thermodynamic limit, so as to defer discussion of boundary conditions.}
In this mapping there is a global $\ZZ$ ambiguity in the sign of the Ising order parameter,  but no additional local ambiguity. When membranes are closed the relevant one-form symmetry is also present microscopically, as we discuss in Sec.~\ref{sec:model} below.

When membranes have holes in them, these mappings break down at the microscopic level. 
But it is natural to expect that if the holes are ``small'',
then in the infrared we will effectively recover a theory of closed membranes, 
in which the above mappings again apply.\footnote{Perturbation theory about the simple closed-membrane limit can also be used to derive an effective longer-ranged Ising Hamiltonian: this is one way to understand the stability of the Ising exponents  \cite{wegner1971duality,fradkin1979phase}.}
Here we translate this heuristic point into a constructive approach.
The key point  is that, whenever the typical size of holes is finite 
(which we define precisely in terms of percolation)
it is possible to \textit{repair} the membranes, configuration by configuration, by  a ``patching'' procedure \cite{somoza2021self}.
This patching procedure is quasi-local, i.e. it involves patches of finite typical size $\ell$.
Crucially, this means that the patching procedure is unambiguous in the thermodynamic limit.
It can also be done efficiently in a Monte Carlo simulation, as described below.

Once the membranes have been patched, we can define the configuration of the fictitious order parameter ${S({\bf r})=\pm 1}$, configuration by configuration (up to a global sign, and taking  boundary conditions into account). 
After patching, we can also define topological string operators without any obstacle.
The Ising correlation function ``$\<S({\bf r})S({\bf r}')\>$'' is really the expectation value of such a string operator. 
The patching procedure allows this correlator to be given a precise meaning and measured in a simulation 
(perhaps even in an experiment), which would be impossible working only with  microscopic observables.

In the 3D $\ZZ$ gauge-Higgs model, the Higgs and confinement lines (shown in Fig.~\ref{Fig:phasediagram}) are completely equivalent, by the exact duality property of the model \cite{wegner1971duality}.
In our convention, the transition line we discuss is the confinement transition. 
As can be seen in Fig.~\ref{Fig:phasediagram} we can cross this line at various values of the Higgs coupling, which is related to $y$ in the figure. 
As we move along the confinement phase transition line by increasing $y$, the typical size $\ell$ of the ``holes'' in the membrane increases, diverging at $y_\text{MCP}$, the self-dual critical point where Higgs and confinement lines meet \cite{somoza2021self, huse1991sponge}.
We study the confinement (Ising$^*$) transition at various values of $y$, up to and including $y_\text{MCP}$.

\tableofcontents

\section{$\ZZ$ worldsheets: review}
\label{sec:model}

We will demonstrate the patching procedure for ``$\ZZ$ membranes'' --- worldsurfaces of $\ZZ$ flux lines in 2+1D. We take the $\ZZ$ gauge field to be coupled to a $\ZZ$ matter field.  We start by reviewing this model. We will be quite brief. See the early parts of Ref.~\cite{somoza2021self} for a more complete overview of the model.

\subsection{Definition of the model}

The partition function, for an ${L\times L\times L}$ cubic lattice with periodic boundary conditions, is
\ba\label{eq:Z3Dgaugerep}
Z \propto 
\sum_{\{\sigma\}, \{\tau\}}
\exp \left[
K \sum_{\square} \lf \prod_{\<ij\> \in \square} \sigma_{ij} \ri
+ 
J 
\sum_{\<ij\>} \tau_i \sigma_{ij} \, \tau_j \right],
\end{align}
where $i$ labels lattice sites, $\sigma_{ij}=\pm 1$ is the gauge field on links, and $\tau_i=\pm 1$ is the Higgs field on sites. We refer to $K$ as the gauge stiffness and to $J$ as the Higgs coupling. However it will be convenient to trade them for the parameterization  \cite{wegner1971duality,balian1974gauge}
\ba\label{eq:definexy}
x & = \tanh K, 
& 
y & = \tanh J
\end{align}
(used in Fig.~\ref{Fig:phasediagram}), which are the natural ``fugacities'' in a membrane representation of the partition function  \cite{huse1991sponge}. 

\begin{figure}
    \centering
    \includegraphics[width=0.8\linewidth]{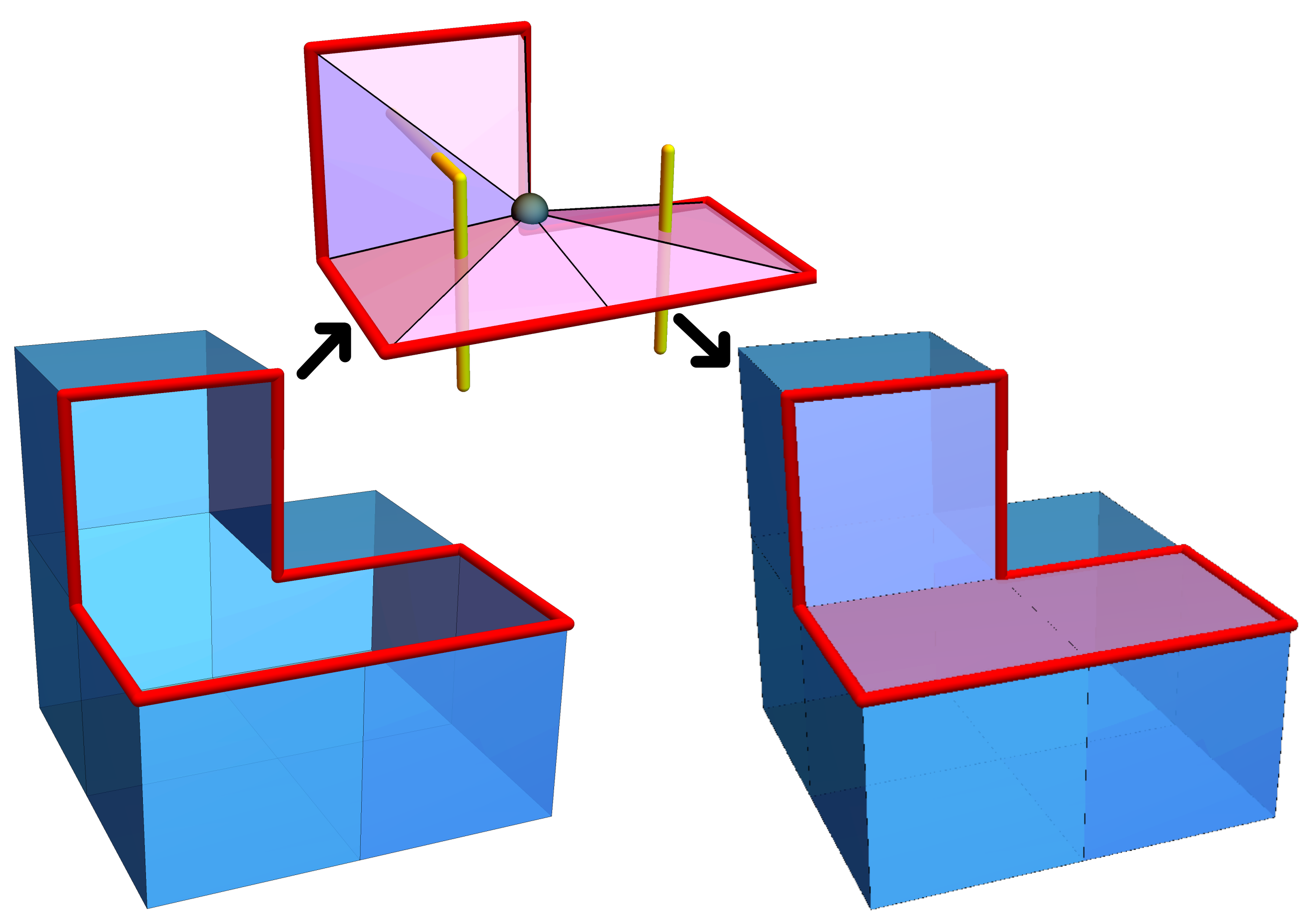}
    \caption{Configuration with a hole  and patching scheme (Sec.~\ref{sec:patchingalgo}).}
    \label{Fig:patching}
\end{figure}

There are two  dual ways to formulate the membrane mapping.
The equivalence of these dual representations is one way to see the equivalence of the Higgs and confinement phase transition  lines in (Fig.~\ref{Fig:phasediagram}). 
In a quantum language,  these dual formulations correspond to formulations of the path integral either in the ``electric'' or the ``magnetic'' basis, as discussed below.
We choose the ``electric'' representation, in which 
the membrane ensemble is\footnote{The electric representation is obtained by a strong coupling (``high temperature'') expansion of Eq.~\ref{eq:Z3Dgaugerep}. 
The dual  (i.e. ``magnetic'') representation is an even more straightforward rewriting of Eq.~\ref{eq:Z3Dgaugerep}:
 the  membranes live on the dual cubic lattice, and a   plaquette of the dual cubic lattice is occupied if $
{\tau_i\sigma_{ij}\tau_j=-1}$ for the 
 corresponding link $\<ij\>$ of the original lattice. See e.g. App.~A of \cite{somoza2021self} for more detail.}
\be\label{eq:partitionfunctionmembranes}
Z = \sum_{\mathcal{M}} \, x^{|\mathcal{M}|} \, y^{|\partial \mathcal{M}|}.
\ee
Here $\mathcal{M}$ stands for a configuration of membranes, which means a set of \textit{plaquettes} of the cubic lattice that are to be regarded as occupied. 
See \cite{somoza2021self} for details.
$|\mathcal{M}|$ is the total membrane area of $\mathcal{M}$.
The membrane boundary $\partial \mathcal{M}$ is a set of occupied \textit{links}: 
A {link} is part of the membrane boundary $\partial\mathcal{M}$ if it is adjacent to an odd number of occupied plaquettes. $|\partial\mathcal{M}|$ is the total length of occupied links.
The left part of Fig.~\ref{Fig:patching} shows an example of a small piece of membrane with a nontrivial boundary.

We will define the term ``hole'' to mean a geometrically connected subset of $\partial\mathcal{M}$, i.e. a cluster of boundary links (see Sec.~\ref{sec:patchingalgo} for a more precise definition). 
It is often useful to idealize holes as \textit{loop-like}, as explained below (Sec.~\ref{sec:patchinglargel}), though microscopically they can have more complicated topology.

(Although for convenience we use the word ``hole'' to denote any
connected component of the membrane boundary $\partial\mathcal{M}$, 
such a boundary component does not need to resemble a hole in a larger piece of membrane:  for example, a ``hole'' could also be the boundary of a  small membrane ``platelet'' made up of a single plaquette.)

From Eq.~\ref{eq:partitionfunctionmembranes} we see that $x$ controls the bare membrane tension, and $y$ controls the bare line tension for holes, i.e. for membrane boundary.

Eqs.~\ref{eq:Z3Dgaugerep} and \ref{eq:partitionfunctionmembranes} can be thought of as 3D classical partition functions in their own right.
However, if we choose to think of $Z$ as an imaginary time path integral for a 2+1D quantum $\ZZ$ gauge theory, then the membrane
ensemble (\ref{eq:partitionfunctionmembranes}) is the formulation of this path integral in the basis of $\mathbb{Z}_2$ electric flux lines which live in the 2D spatial plane and which terminate on electric charges.
The membranes we are discussing are  worldsurfaces of these electric flux lines (see e.g. Fig.~6 in Ref.~\cite{somoza2021self}). 
The boundaries of the membranes, which make up the ``holes'', are worldlines of the electric charges.
Duality is the existence of a completely equivalent picture in the magnetic basis, for $\ZZ$ magnetic flux lines and charges.

Fig.~\ref{Fig:phasediagram} above showed the phase diagram. The key features are the confinement and Higgs transition lines (related by duality); the self-dual multicritical point  at ${(x_\text{MCP}, y_\text{MCP})}$, where confinement and Higgs lines both end \cite{fradkin1979phase,TupitsynTopological,vidal2009low,somoza2021self}; and a short first-order segment to the left of the multicritical point.

We will focus particularly on the confinement transition line, marked in solid red in Fig.~\ref{Fig:phasediagram}. 
If we move along a \textit{horizontal} line on the phase diagram at fixed $y$ in the range ${[0,y_\text{MCP})}$,
then we cross this line at an $x$-value that we will denote $x_c(y)$.
Loosely speaking, 
crossing this line from the left to the right gives a  proliferation of membranes 
\cite{huse1991sponge}.

A key point will be that if we move upwards along  a \textit{vertical} line in the phase diagram, at any  fixed $x$, there is a critical value of $y$  
where the characteristic size $\ell$ of holes\footnote{One way to  define $\ell(x,y)$ at a given point $(x,y)$ in the phase diagram is via the   exponential decay constant for the probability that a given region of size $R$ contains a hole of linear size $R$: $\text{prob}\sim e^{-R/\ell}$.} diverges  \cite{huse1991sponge, somoza2021self}. We will refer to this as the ``hole percolation'' threshold.  
The location of the transition line is shown in Fig.~\ref{Fig:phasediagramwithpercolation} below.
This threshold  coincides with the Higgs  transition for ${x>x_\text{MCP}}$, and passes through the multicritical point \cite{somoza2021self}.
The percolation transition line also continues to ${x<x_\text{MCP}}$, but its nature changes in this regime  and it need no longer have any thermodynamic significance for ${x<x_\text{MCP}}$.

\subsection{The ${y=0}$ limit: closed membranes}
\label{sec:vanishingylimit}

The lower boundary of the phase diagram, 
i.e. the line $y=0$,
is special:
the partition function simplifies there because the membranes are closed   ({$\partial\mathcal{M}=0$}).

This means that on the line $y=0$ there is a microscopic mapping to Ising \cite{wegner1971duality}, and that topological line operators (t'Hooft loops) can be defined microscopically 
\cite{wegner1971duality,fradkin1979phase,nussinov2009symmetry,kapustin2017higher,gaiotto2015generalized,kitaev2003fault}.
We now review how this works for $y=0$, in preparation for our main aim which is to extend these mappings and definitions to ${y>0}$.

The closed membranes at $y=0$ can be mapped to Ising domain walls, by introducing  Ising spins $S_r$ that live at the centres of cubes.
(That is, $r$ is a site of the dual cubic lattice, made up of cube centres.
A pair of adjacent spins $S_r$, $S_{r'}$ are antiparallel if the corresponding plaquette $p$ of the original lattice is occupied, i.e. $p \in \mathcal{M}$.)
The Ising spin $S$ is ordered for ${x<x_c{(0)}}$, i.e. in the confined phase, 
and $S$ is disordered for ${x>x_c{(0)}}$, i.e.
in  the deconfined phase, where membranes proliferate.

In the finite system, the mapping is to an Ising model with a sum over both periodic and antiperiodic boundary conditions in each direction:
this is because  a closed path that loops around the system in, say, the $x$ direction can cross either an even or an odd number of membranes.
Denoting these two possibilities by ${\sigma_x= \pm}$ and similarly for the $y$ and $z$ directions, the gauge theory partition function decomposes into sectors:
\be
\label{eq:Z_sectors}
Z = \sum_{\sigma_x, \sigma_y, \sigma_z = \pm} Z_{\sigma_x, \sigma_y, \sigma_z}.
\ee
Finally, the relation to the Ising partition function involves a factor of 2, because the above definition of the Ising spin $S_r$ has a global sign ambiguity: 
${Z_{\sigma_x, \sigma_y, \sigma_z}^\text{Ising}=2Z_{\sigma_x, \sigma_y, \sigma_z}}$, where on the LHS $\sigma_\mu=\pm$ denotes periodic/antiperiodic Ising BCs in the $\mu$ direction.

While there is a ``duality'' between the gauge theory at ${y=0}$ and the Ising model,\footnote{This ``duality'' between the gauge theory and the Ising model should not be confused with the self-duality of the gauge theory that relates the electric and magnetic representations.}  
the Ising spin $S_r$ does not exist as an observable in the original gauge theory.
However the gauge theory does have well-defined string operators $V_P$ 
(observables defined on a line)
that are closely related to correlators such as ${\<S_r S_{r'}\>}$. 

Given a path $P$ on the dual lattice (either closed or open), 
$V_P$ is equal to ${+1}$ if $P$ intersects an even number of occupied plaquettes and to ${-1}$ if $P$ intersects an odd number of occupied plaquettes. (We can write $V_P$ as a product of local signs for each plaquette pierced by $P$.)
Because all membranes are closed when $y=0$,
the value of $V_P$ does not change if we deform the path $P$ (keeping its endpoints, if any, fixed).

If $P$ is an open path between $r$ and $r'$,  in an infinite system,  then $V_P$
depends only on its endpoints. From the domain wall interpretation, it is clear that under the mapping to Ising
\be
V_P \longleftrightarrow S_r S_{r'}.
\ee 
In  the finite system with periodic boundary conditions $V_P$ is also nontrivial for closed loops that wrap around the system. ($V_P$ is trivial for a contractible loop.)

The operators $V_P$ can be viewed as symmetry operators \cite{nussinov2009symmetry,kapustin2017higher,gaiotto2015generalized}.
For example, in the Hamiltonian formulation of the gauge theory, as a 2D quantum system, one may define corresponding unitary quantum operators $\hat V_P$  for paths $P$ lying in the spatial plane. The invariance of $V_P$ under deformations of $P$ implies that these operators  commute with the Hamiltonian: ${[\hat H, \hat V_P]=0}$. 
Unlike the generators of a conventional symmetry (which would be supported on the entire spatial plane) these operators are supported on one-dimensional lines. 
They are referred to as 1-form symmetry operators. 

In this language, the one-form symmetry is spontaneously broken in the deconfined phase, because of the existence of a degenerate ground state manifold within which $\hat V_P$, for winding loops, acts nontrivially. 
In the language of the $\ZZ$ topological order and the toric code, $\hat V_P$ are the ``magnetic'' string operators associated with ``m'' excitations \cite{kitaev2003fault}.

\subsection{Finite and infinite holes}

The above concepts (the ``dual'' Landau order parameter $S_r$, and the string operators) are useful for detecting the deconfined phase and the transition out of it. 
To generalize them
to ${y>0}$, we will need to deal with the problem of having holes in the membranes.
We make some qualitative points before giving a concrete construction in the next Section.
For concreteness we focus the discussion below on the confinement transition, but the key points also apply in the deconfined phase.
By duality, the discussion also translates directly to the Higgs transition. We will discuss some other extensions of the method in Sec.~\ref{sec:extensions}.

\begin{figure}[t]
    \centering
\includegraphics[width=0.96\linewidth]{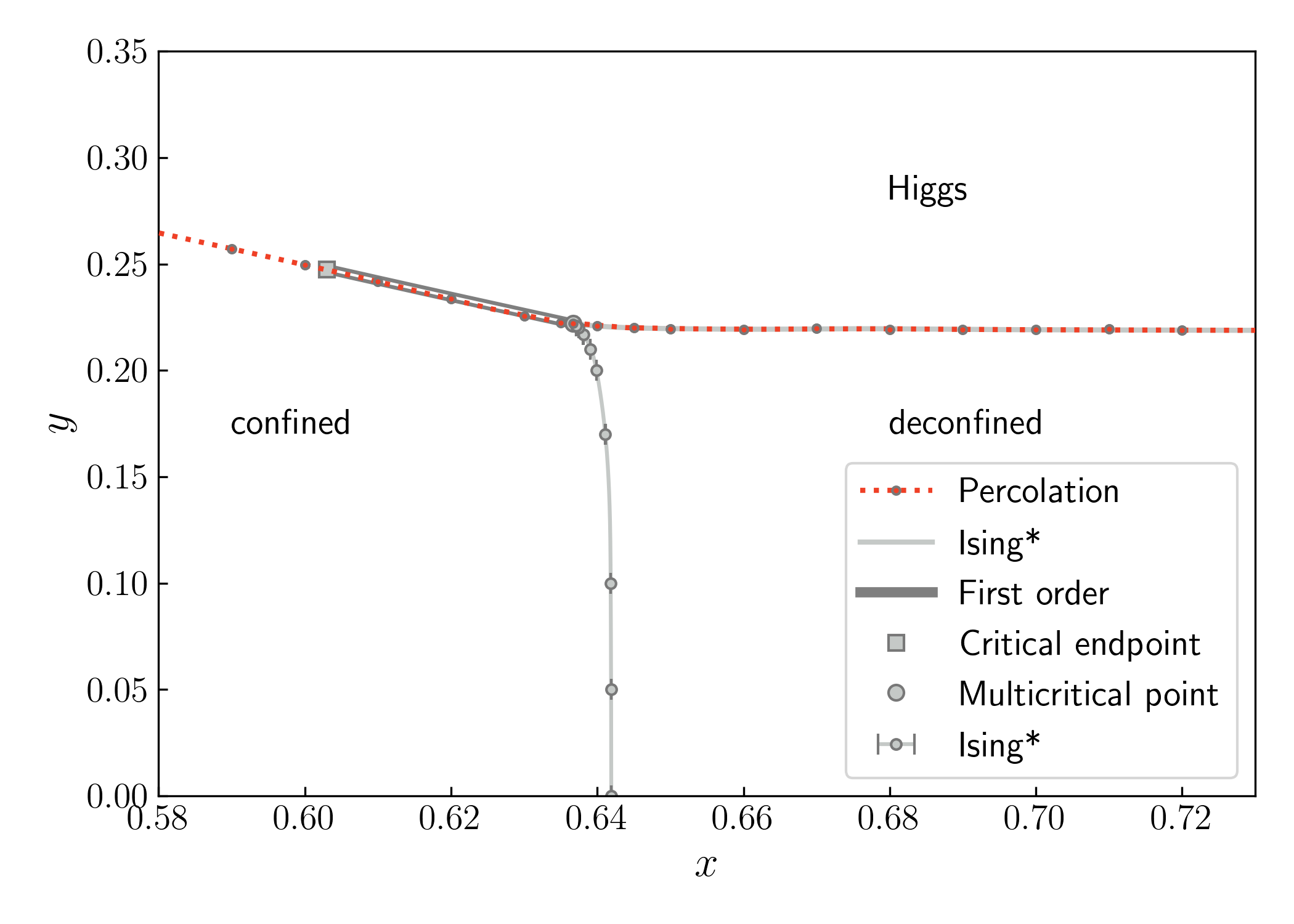}
    \caption{ The  phase diagram of Fig.~\ref{Fig:phasediagram} shown
    together with   the percolation threshold   for holes 
    (red dashed line).  The typical hole size $\ell(x,y)$ is finite below this line (data from \cite{somoza2021self}).}
\label{Fig:phasediagramwithpercolation}
\end{figure}

Let ${\ell(y) \equiv \ell\big(x_c(y),y\big)}$ be the typical hole size at a given point along the confinement transition line. 
$\ell(y)$ grows with increasing $y$, and diverges at the MCP.
Fig.~\ref{Fig:phasediagramwithpercolation} shows the part of the phase diagram where $\ell$ is finite.

Heuristically,  when $\ell(y)$ is finite (${y<y_\text{MCP}}$), we might imagine that holes disappear under RG. 
That is, after coarse-graining
to scales ${\gg \ell(y)}$
we effectively recover a theory of closed membranes, 
and the system  flows to the same Ising$^*$ fixed point that governs the phase transition on the boundary ${y=0}$. 
Below we will show that this can be made concrete (giving a constructive mapping to Ising)  also for ${y>0}$.

We construct the  dual spin $S_r$ explicitly and show that it  only behaves like a local operator on scales much larger than $\ell(y)$.
Therefore  the duality to Ising only \textit{emerges} beyond this scale.

Similarly  string operators analogous to  $V_P$ can  be defined as ``fattened'' objects \cite{hastings2005quasiadiabatic}. Because of rare large loops, the string operators must in general be much thicker than $\ell(y)$, as we discuss below.

We will also examine  the fate of these concepts at the multicritical point (MCP), where $\ell=\infty$
and ``holes'' are present on all scales.
We emphasise that the MCP is in a different universality class to the confinement line. Monte Carlo strongly indicates that the multicritical point is a scale-invariant critical point, which is of interest because it is a simple example of a critical point without any known useful continuum Lagrangian description~\cite{somoza2021self}.

The dual order parameter $S_r$ ceases to be local at the multicritical point (MCP). 
This is one way of understanding why the 
mapping to a Landau theory breaks down at the MCP (see   Refs.~\cite{somoza2021self,oppenheim2023machine} and Sec.~\ref{sec:outlook} for further discussion of this point).

Interestingly, though, there is a sense (discussed in Sec.~\ref{sec:selfdual} below) in which this nonlocality is mild at the MCP.  This may be a hint as to why various exponents at the multicritical point are numerically close to exponents of the XY model \cite{somoza2021self,bonati2022multicritical}.

\section{Membrane patching}

\subsection{Schematic}
\label{sec:patchingschematic}

To define the dual order parameter and the string operators,  we will   define a convention for mapping a given membrane configuration $\mathcal{M}$ --- which in general will have a nontrivial boundary, ${\partial\mathcal{M}\neq 0}$ --- to a ``repaired'' or ``patched'' configuration
$\overline{\mathcal{M}}$ with 
${
\partial \overline{\mathcal{M}}=0
}$.
Once this is done,
$\overline{\mathcal{M}}$ can be viewed as a configuration of domain walls for a dual spin $S_r$, and 
observables can be constructed in terms of $\overline{\mathcal{M}}$ in the same way as at ${y=0}$.

We emphasize that this is a procedure for defining \textit{observables} in a given configuration $\mathcal{M}$ --- 
we leave the partition function unchanged.
In particular, above we defined string operators $V_P(\mathcal{M})$ in the closed membrane ensemble (here we have made the  dependence of this observable on  the configuration $\mathcal{M}$, as well as on a path $P$, explicit).
The patching procedure allows us to define a string operator $\widetilde{V}_P$ in the ${y>0}$ ensemble, via:
\be
{\widetilde{V}_P(\mathcal{M}) \equiv V_P(\overline{\mathcal{M}})}.
\ee
Since $\overline{\mathcal{M}}$ is closed, $\widetilde{V}_P$ has the key invariance property under deformations of the path $P$.
For an open path $P$ between points $r$ and $r'$
in an infinite system
we have the duality relation
${\widetilde V_P(\mathcal{M})
\leftrightarrow S_r S_{r'}}$ 
(we discuss boundary conditions in a finite system below).

We will perform the repair by attaching a patch for each hole separately
(by the rule in the following subsection).
Note that the mapping from $\mathcal{M}$ to $\overline{\mathcal{M}}$ is not  a strictly local operation: 
a given plaquette in 
$\overline{\mathcal{M}}$  depends on other plaquettes in $\mathcal{M}$. 
In turn, this relaxes the locality properties of $\widetilde{V}_P$ and the dual spin $S_r$.
However, if the typical hole size is finite, there is an effective notion of  locality on larger scales \cite{somoza2021self}, which we discuss in Sec.~\ref{sec:locality}.

We now describe concretely how to define the patching.
In following sections we will implement it
in a Monte Carlo simulation of Eq.~\ref{eq:partitionfunctionmembranes},
focussing particularly on the confinement transition for ${0<y<y_{MCP}}$.

\subsection{Membrane-patching algorithm}
\label{sec:patchingalgo}

The membrane-patching algorithm
is as follows.
For a given configuration $\mathcal{M}$, the first step consists in performing a percolation analysis of the occupied links (links in ${\partial \mathcal{M}}$).
This means that we determine the   connected clusters of links. 
(A cluster does not share any site with any other cluster.)

If there is a cluster that spans the entire system in any direction\footnote{A cluster is defined to span the system in a given direction, say the $X$ direction, if for every $X$-coordinate value $x\in\{0,\ldots,L-1\}$
the cluster contains at least one link occupying it.}
then we flag the configuration as percolating (``non-patchable''), and do not try to patch it.
Later when we compute expectation values, 
we will often condition on the configuration being patchable (non-percolating).
When ${y<y_\text{MCP}}$, the probability of a configuration percolating is vanishingly small at large $L$.

In a given patchable configuration,
we patch each  cluster separately.
The links of a cluster represent the boundary of a patching surface: see for example the left image in Fig.~\ref{Fig:patching}. 
It is convenient to think of the definition of this surface as a two-step procedure.

\begin{figure}
    \centering
    \includegraphics[width=0.7\linewidth]{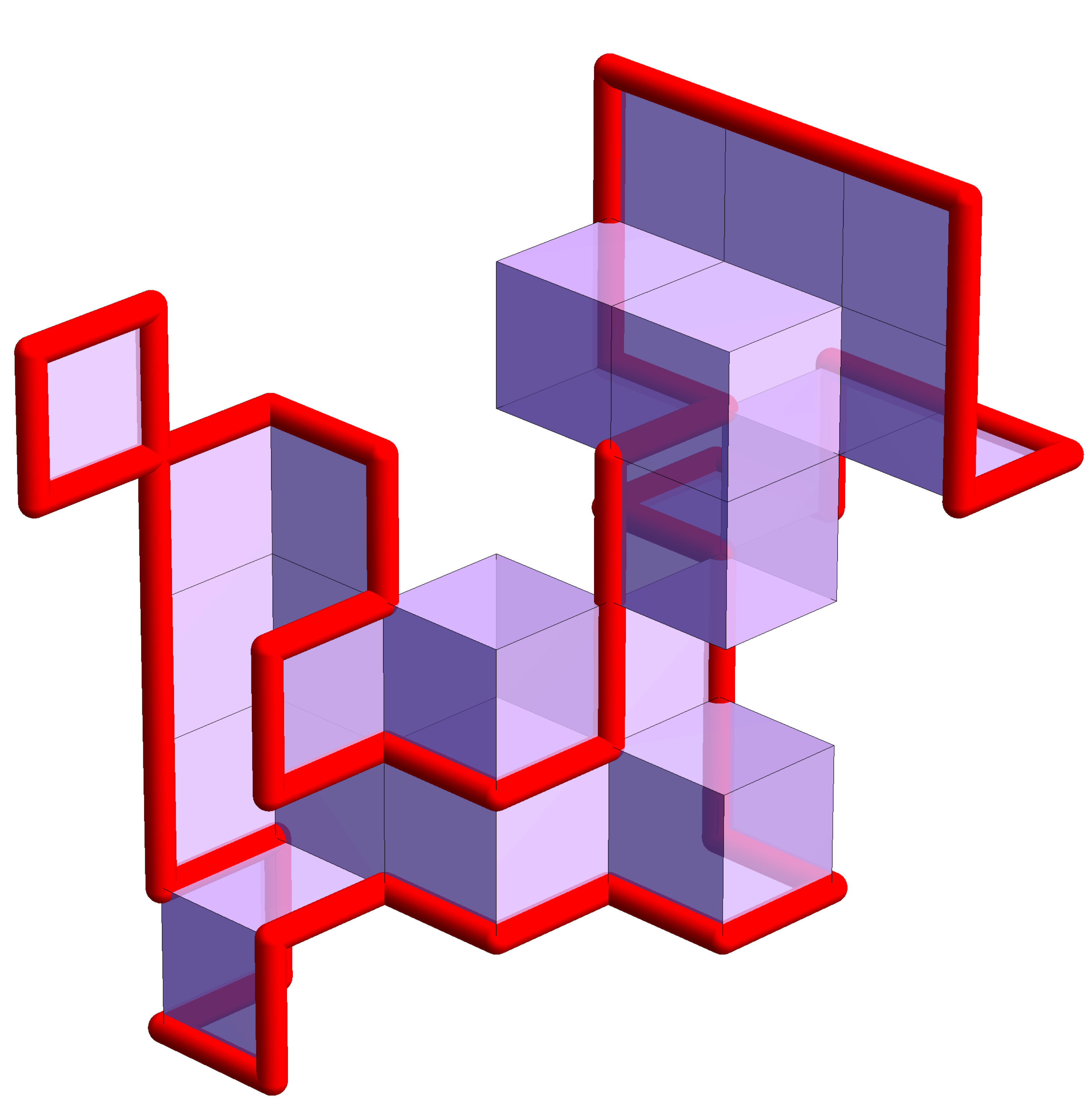}
    \caption{A loop configuration (``hole'') and the resulting membrane patch produced by the algorithm.}
    \label{Fig:loopandpatch}
\end{figure}

First, a patching surface can  easily be constructed as a union of triangles. For each link in the cluster, we  define a triangle with the link as one side and the cluster's center of mass as the opposite vertex. 
Such a surface is shown in the central image in Fig. \ref{Fig:patching}. 

Next, we simplify this surface to an equivalent 
one  constructed from plaquettes of the cubic lattice, instead of from triangles (Fig. \ref{Fig:patching}, Right).
Note that each plaquette $p$ is dual to a link $l(p)$ of the dual cubic lattice
(lines in yellow in the central image in Fig.~\ref{Fig:patching}).
Loosely speaking, if a triangle crosses a dual link $l(p)$, this means that the plaquette $p$ should be included in the patch.
More precisely, a plaquette $p$ is included if the dual link $l(p)$   crosses an odd number of triangles. An example of the membrane patch associated to a cluster is shown in Fig.~\ref{Fig:loopandpatch}.

A final detail is that in rare cases a triangle may pass exactly through a vertex of the dual cubic lattice.
For a simple solution, we resolve the degeneracy by infinitesimally perturbing the centre of mass in a random direction. Other resolutions are possible, but the choice will not have any significant effect on the results.

We have proposed an explicit, easy to implement, patching procedure. Other legitimate procedures,
leading to slightly different patched surfaces,
are also possible, but such changes would not affect the
universal behavior of the 
correlation function at distances $r\gg \ell$ (Sec.~\ref{sec:diagnosingdeconf}, Sec.~\ref{sec:locality}), or the analysis of the phase transition.
Having defined the patches we can define the repaired configuration $\overline{\mathcal{M}}$.
If we call the set of patching plaquettes ${\cal P}$, 
the new set $\overline{\mathcal{M}}={\cal M}\cup {\cal P}-{\cal M}\cap {\cal P}$  defines strictly closed surfaces and allows us to construct line operators as described in Sec.~\ref{sec:patchingschematic}.

\subsection{Ising observables}
\label{sec:Isingobservables}

In particular, we can define correlators of the dual Ising field, e.g. ${\<S_r S_{r'}\>}$, if we correctly account for boundary conditions (as discussed directly below).
We propose that these correlators are practically useful for the  analysis of the phase transition, because they allow standard tools for \textit{order parameters}   to be applied to the  confinement and Higgs transitions, despite the fact that the original model does not have an order parameter.
For example, such tools include the analysis of 
the total magnetization, and of Binder cumulants, to accurately identify the transition point.
(We will use the language of the confinement transition, but by duality,  the results translate directly to the Higgs transition.)
We reiterate that for $y>0$ the quantity $S_rS_{r'}$ is definable only through the  patching procedure, unlike the simpler case $y=0$ where $S_rS_{r'}$ is the expectation value of a ``microscopic'' line operator.

As discussed around Eq.~\ref{eq:Z_sectors}, an ensemble of  closed membranes can be split into sectors. 
Our patching algorithm works  in all sectors.
It is also perfectly meaningful to define correlation functions  of the $\widetilde V$ operators conditioned on a given sector (as defined more precisely below):
correlators for different sectors agree when $r\ll L/2$, so that restricting to a sector does not affect the  location of the phase transition,
but may differ when $r$ is of order $L$. 

One reason it is often useful to restrict to a sector is for the numerical analysis of the confinement phase transition. 
The analysis of the transition  through the asymptotics of the correlators is subject to large statistical errors; more effective procedures are based on the analysis of Binder-like parameters \cite{binder1981critical}.
These can be defined straighforwardly in the first sector
(${\sigma_x=\sigma_y=\sigma_z=+}$). 

In this sector the  dual Ising spin $S_r$ has periodic boundary conditions, and so is well-defined up to a global sign flip: this 
means that, after patching, the correlation function $\<S_r S_{r'}\>$ may be straightforwardly defined. When we write expressions such as $\<S_r S_{r'}\>$, we refer to an average conditioned on the 
${\sigma_x=\sigma_y=\sigma_z=+}$ sector.

More formally, this may be expressed as 
\be
\<S_r S_{r'}\> \equiv
\f{\< \widetilde V_P(\mathcal{M}) \, \chi_{+}(\mathcal{M}) \>}{\<\chi_{+}(\mathcal{M})\>},
\ee
where $P$ is any path connecting $r$ to $r'$, and $\chi_{+}(\mathcal{M})$ is an indicator function that takes the value $\chi_{+}=1$  if $\mathcal{M}$ is a patchable configuration in the $+++$ sector, and the value $\chi_{+}=0$ otherwise.

Similarly,
for a configuration in the ${\sigma_x=\sigma_y=\sigma_z=+}$ sector,
the squared magnetization density
\be\label{eq:defnmagnetization}
M^2 = \Big( L^{-3}\, {\sum}_r S_r \Big)^2 
\ee
is well-defined. 

In the simulations below it will also be convenient to define two dimensionless quantities: 
\begin{align}
\operatorname{b}_4(M)&= \frac{3 \< M^2\>^2-\< M^4\>}{2\<M^2\>^2}, \\
 \operatorname{b}_1(M)&=\frac{1}{1-\sqrt{2/\pi}}\left(\frac{\<|M|\>}{\<M^2\>^{1/2}}
    - \sqrt{\frac{2}{\pi}} \,
    \right).
    \label{eq:b1a}
\end{align}
$\operatorname{b}_4(M)$ is just the Binder parameter \cite{binder1981critical} rescaled such that it becomes 1 in the ferromagnetic phase and 0 in the paramagnetic one ($\operatorname{b}_4=3 U_L/2$). $\operatorname{b}_1(M)$ behaves similarly to $\operatorname{b}_4(M)$ and has the same two limits,  but it is statistically slightly easier to estimate \cite{somoza2021self},  so we suggest the use of the former instead of the latter.

We will also store statistics related to the total patched area, which we denote $A_\text{patched}$:
for a given configuration,
$A_\text{patched}$  is the number of plaquettes in the set $\mathcal{P}$ defined above in Sec.~\ref{sec:patchingalgo}.
Analysis of the statistics of 
$A_\text{patched}$
at different points on the confinement line
reveals the divergence of the patching lengthscale $\ell(y)$
as the self-dual critical point is approached.

\begin{figure*}[t]
    \centering
    \includegraphics[width=1.\linewidth]{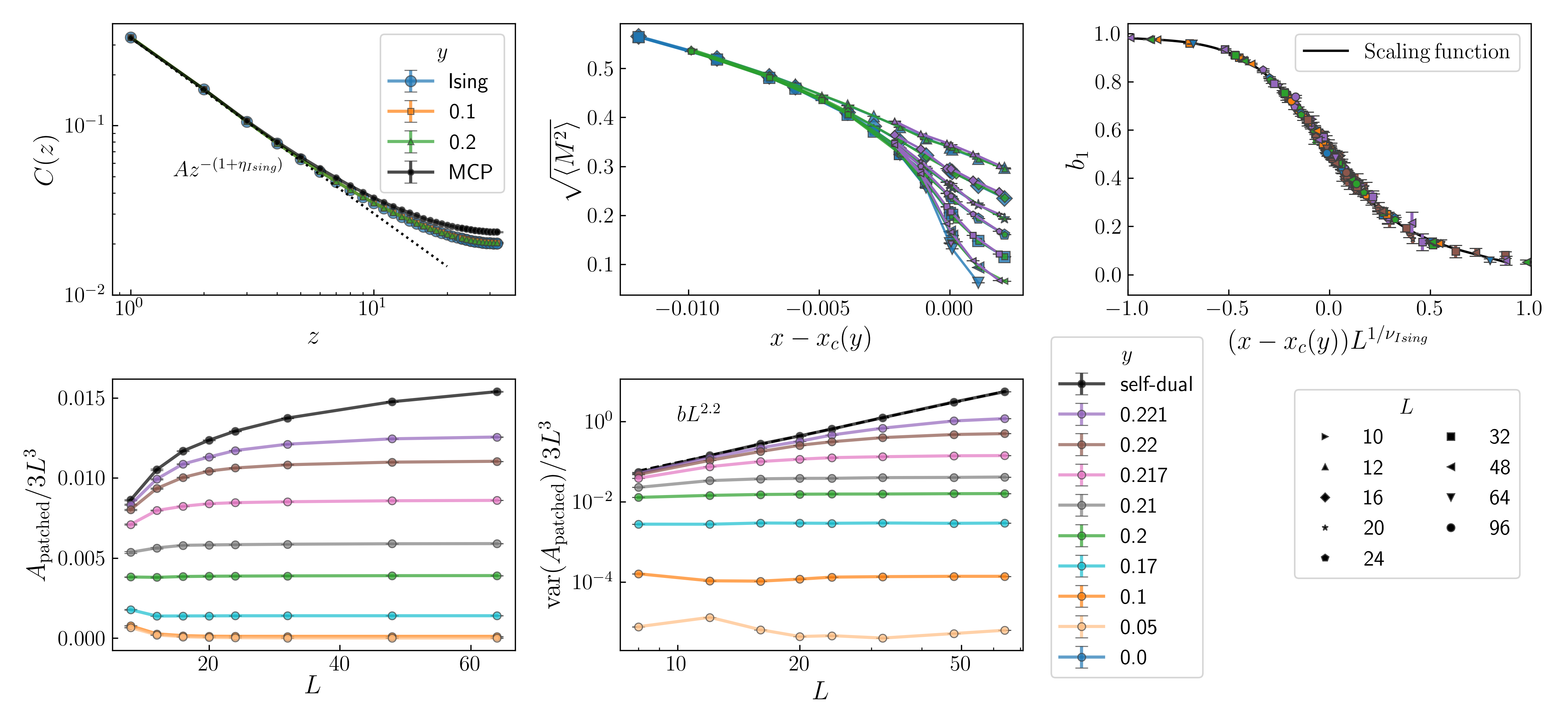}    \caption{ Panel 1 -- Correlation 
     as a function of  distance
    along one of the axis directions 
    for system size $L=64$ at the critical point $x_c(y)$ for several values of $y$. Black dotted line represents a power-law $r^{-(1+\eta_\mathrm{Ising})}$, where $\eta_{Ising}=0.036298$.  Panel 2 -- Magnetization as a function of ${x-x_c(y)}$, for several values of $y$ and different system sizes. Colours and symbols indicate $y$ values (three are plotted) and system sizes, respectively.  
    Panel 3 -- Scaling collapse of parameter $b_1$ as a function of ${(x-x_c(y))L^{1/\nu_\mathrm{Ising}}}$ where $\nu_\mathrm{Ising}=0.629971$ is the critical exponent of the 3D Ising universality class. Panel 4 -- Patched area for several values of $y$ as a function of $L$ at their critical point $x_c(y)$. Panel 5 -- Variance of the patched area for several values of $y$ as a function of $L$ in logarithmic scale,  at their critical point $x_c(y)$. 
    }
    \label{fig:6panels}
\end{figure*}

\subsection{
Diagnosing the deconfined phase; \\
Robustness of patching procedure
}

\label{sec:diagnosingdeconf}

Since in Sec.~\ref{sec:resultsisingstar} below we will focus  on critical scaling near the confinement transition line (the red line in Fig.~\ref{Fig:phasediagram}), 
we emphasize in this section that the patching procedure  gives a way to detect or verify deconfinement more generally (even far away from the confinement transition line).

First we note that the conditions ($i$)  and ($ii$)  below serve as (easily checkable) \textit{sufficient} conditions for a point of the phase diagram to lie within the deconfined phase. 
Subsequently we will argue that, for the present patching algorithm and model, they are also \textit{necessary} conditions (meaning that patching can be used as a diagnostic of deconfinement throughout the entire deconfined phase):

$\bullet$ Condition ($i$): The patching procedure succeeds: 
i.e. the probability of a configuration $\mathcal{M}$ being patchable tends to 1 at large system size. 
Equivalently, the holes are \textit{non-percolating}, as discussed below.

This condition  ensures that the line operators $\widetilde V$ that we define through patching are quasilocal (as discussed in more detail in Sec.~\ref{sec:locality}).
Given this, these operators may then be used to detect deconfinement.

In principle, one way to detect deconfinement  would be  by checking for the spontaneous breaking of one-form symmetry using  a line operator that wrapped around the system in some direction.
An alternative (and more practical\footnote{In the quantum language, a signature of the deconfined phase is that a line operator $\hat{\widetilde{V}}$, which  winds around the system in some direction, acts  nontrivially on the ground state space.
As a result $\langle \widetilde V \rangle =0$ for the thermal expectation value of a winding string operator, 
so long as the thermal expectation value properly sums over all topological sectors (Sec.~\ref{sec:vanishingylimit}).
However, because of the order--$L$  free energy barrier that separates different topological sectors (in the 3D classical language), it will often be much easier for Monte Carlo to sample a single sector than to sample all sectors. Therefore observables that can be calculated in a single sector are more useful in practise. This is not a serious restriction since it is possible to detect the thermodynamic phase using observables within a topological sector.}) diagnostic is the expectation value of an open string operator:

$\bullet$ Condition ($ii$): The expectation value of the string operator ${\lim_{L\to\infty} \langle\widetilde V_P\rangle}$, 
for an open path $P$ of length $R$, tends to zero at large $R$.

In terms of the dual Ising degrees of freedom, this means that the dual Ising order parameter is \textit{disordered}. 
Numerically, a convenient way to detect this is via the vanishing at large $L$ of the squared magnetization $\langle M^2\rangle$ (defined in the previous section).

The two conditions above show that patching  gives a simple diagnostic for the deconfined phase: first we confirm that holes do not percolate, and then we check that $\langle M^2\rangle$ vanishes at large $L$.

If, in the second step, we find that $\langle M^2\rangle$ does not vanish, this tells us that
 we are not in the deconfined phase. This is what happens if we are to the left of the confinement line in Fig.~\ref{Fig:phasediagram}.

The remaining question is about the nature of the phase transitions where the holes start to percolate, 
i.e. where condition ($i$) ceases to hold.
 A priori, there are two possibilities.
First, the percolation transition of the holes could happen at the  thermodynamic phase boundary of the deconfined phase (the Higgs transition line in Fig.~\ref{Fig:phasediagram}, together with the multicritical point).
Alternately,  the percolation transition line could  be separated from the thermodynamic phase transition line, occurring at a smaller value of $y$ that had no thermodynamic significance. 

For the present model, the former scenario holds.
In other words, the Higgs transition line \textit{is} the percolation phase transition of the holes.
This was already anticipated in Ref.~\cite{huse1991sponge};
Ref.~\cite{somoza2021self} gave numerical evidence that the hole percolation transition coincides with the thermodynamic transition along the full extent of the Higgs line (plus multicritical point),
and sketched an analytical explanation for why the two transitions coincide which we develop a little further in App.~\ref{sec:percolationanalytic}.

As a result, the present patching approach  successfully diagnoses the entirety of the deconfined phase.  Further, we argue in App.~\ref{sec:percolationanalytic} that this property is robust to slight changes of the model or the patching scheme.

On the other hand, a  sufficiently severe change to the way clusters are defined in the patching scheme
(a sufficiently ``bad choice'' of patching algorithm)
could cause percolation 
to occur in the interior of the deconfined phase, as we discuss in App.~\ref{sec:percolationanalytic}.
In that scenario, the ``bad'' patching algorithm would still be able to successfully diagnose deconfinement for small enough $y$, 
but for larger $y$ it would not give us any information about which phase the model was in.

\subsection{Monte Carlo updates}
\label{sec:updates} 

For completeness we specify the updates used to equilibrate the model. For one Monte Carlo step we use two standard Metropolis updates: First we attempt to update all plaquettes in the system (one at at a time). 
Second, for all  cubes in the system, we try to change all six plaquettes of the cube at the same time. 
For ${y=0}$ this second procedure reproduces the standard spin-flip update for the dual Ising system.

\section{Application to Ising$^*$ transition}
\label{sec:resultsisingstar}

We now apply the above algorithm to the confinement phase transition (${y<y_\text{MCP}}$).
By the exact self-duality of the gauge theory, the following results also apply to the Higgs transition, in the dual membrane representation (where the membranes are worldsheets of  magnetic, rather than electric,  fluxlines).

First we show results for dual Ising observables which are computable                   using this patching algorithm.
Then we study the properties of the patches themselves: 
these reveal a growing lengthscale $\ell(y)$, and a fractal structure below this scale, when $y$ gets close to $y_\text{MCP}$.

\subsection{Dual Ising correlators}

We study the phase transition at several values of $y$.
For a given value of $y$, the critical value $x_c(y)$ can be determined using the modified Binder cumulant $b_1$ (Eq.~\ref{eq:b1a}), which shows a crossing at $x_c(y)$, and is discussed further below.

A striking demonstration of Ising$^*$ universality is given by the power-law scaling of the dual Ising correlator as a function of separation $z$:
\be
C(z) = \langle S_{(0,0,0)}S_{(0,0,z)}\rangle
\ee
at the critical point ${x=x_c(y)}$.
At $y=0$ this correlator can be defined microscopically using a simple line operator,
but for $y>0$ it is nontrivial and must be defined using the patching procedure.

The  correlator is shown in Panel 1 of Fig.~\ref{fig:6panels}, as a function of distance $z$ in a system of fixed size $L=64$. 
Data for the values ${y=0,0.1,0.2}$, which lie on the confinement line and are expected to be in the Ising$^*$ universality class, are consistent with the Ising power law, where $\eta_\mathrm{Ising}=0.036298(2)$ \cite{kos2016precision}, which is shown with a dotted line.\footnote{The power law applies for $1\ll r\ll L$. It is also possible to perform a scaling collapse for $r/L$ of order 1 (not shown).} 
(Since the Ising anomalous dimension $\eta_\text{Ising}$ is small, this line is close to $1/z$.)
The distinct curves are also consistent with each other within error bars:
although in general the nonuniversal normalization of
$C(z)$ will depend on $y$, this dependence appears to be very weak.
The Panel also shows the correlator computed at the MCP, where its RG interpretation is different and will be discussed in Sec.~\ref{sec:selfdual}.

An intuitive way to see the confinement transition
is in the dual magnetization (Eq.~\ref{eq:defnmagnetization}), shown
as a function of ${x-x_c(y)}$
in 
Panel 2 of 
Fig.~\ref{fig:6panels}.
Data is shown for different systems sizes and also different $y$.
Again, $y=0$ is the case where there is a microscopic mapping to Ising. 
This is compared with the same quantity for  ${y = 0.2}$ and ${y=0.221}$. 
They both show a striking similarity to the ${y=0}$ case.
Note that we have absorbed the $y$-dependent shift to the location of the critical point (which can be seen in Fig.~\ref{Fig:phasediagram}) so that the transition occurs at the same value of the horizontal coordinate in each case.

Finally, Panel 3 of Fig.~\ref{fig:6panels} shows the scaling collapse of $b_1$ for several values of $y<y_{MCP}$ using the known Ising correlation length exponent  $\nu_\mathrm{Ising}=0.629971(4)$ \cite{kos2016precision}.  
We fitted all data to a single scaling function,
shown in the figure as a black line.  For the scaling function we used 8 free paramaters (8 coefficients for a B-spline function). Additionally, in order to achieve collapse of the curves we kept $x_c(y)$ as an adjustable parameter for each y value used.

Using  data for $y \le 0.217$ and dropping the smaller system sizes we obtained $\chi^2=42.04$ for 39 degrees of freedom, showing an excellent  overlap  between different locations on the phase transition line.
We also obtained accurate phase boundaries $x_c(y)$ from the fit.
These agree within error bars with the estimates 
using percolation observables in \cite{somoza2021self} (the latter estimated the Higgs phase boundary, but duality relates it to the confinement boundary under discussion here).
In the panel we also show data for smaller sizes, as well as data for $y=0.22,0.221$ that were not used for fitting, and the deviation from the scaling curve is still quite small.

\subsection{Other sectors}

The observables above are expectation values conditioned on being in the first sector (${\sigma_x=\sigma_y=\sigma_z=+}$) of the membrane ensemble.
It is of course possible to consider observables averaged over sectors.

A simple one is the probability $P_{\sigma_x,\sigma_y,\sigma_z}$ 
of being in a given sector. 
Note that the value of $\sigma_x=\pm 1$ is just the  value of the string operator $\widetilde{V}_{P_x}$ for a path $P_x$ that winds around the sample in the $x$-direction, so $P_{\sigma_x,\sigma_y,\sigma_z}$  is simply related to expectation values of such winding string operators.

At the Ising$^*$ critical point, and in the limit $L\rightarrow\infty$, these numbers are universal and related by duality to ratios of Ising partition functions with different boundary conditions,
\be
P_{\sigma_x,\sigma_y,\sigma_z} = \f{Z^\text{Ising}_{\sigma_x,\sigma_y,\sigma_z}}{
\sum_{\sigma_x',\sigma_y',\sigma_z'} Z^\text{Ising}_{\sigma_x',\sigma_y',\sigma_z'}
}.
\ee
In a finite-size system, there is also a nonzero probability $P_\text{perc}(L)$ of being in a percolating (non-patchable) configuration, which is not assigned to any of the above sectors. For ${y<y_\text{MCP}}$ this probability vanishes exponentially at large $L$, so that (asymptotically) every configuration can be assigned to an Ising sector.

For ${y=y_\text{MCP}}$, where ${\ell=\infty}$,
the limit $\lim_{L\rightarrow\infty}P_\text{perc}(L)$  is instead a nonzero constant, which we estimate in Sec.~\ref{sec:selfdual} (Table~\ref{tab:sectorsMCP}).

In our Monte Carlo dynamics, the configuration can make a transition between different sectors by passing through a non-patchable configuration with a percolating hole. 
In principle this allows equilibration over all sectors. 
In practice this is possible if $y$ is not too small and $L$ is not too large.
When $y<y_\text{MCP}$ such transitions are  exponentially suppressed in the limit $L\rightarrow \infty$
simply because  $P_\text{perc}(L)$ is exponentially suppressed. 
For this reason we have not tried to accurately compute the sector probabilities for the  Ising$^*$ critical point.
This could be done, using the present gauge theory, by parallel tempering, exchanging configurations at different values of~$y$~\cite{katzgraber2009introduction}.

\subsection{Properties of the patching for large $\ell$}
\label{sec:patchinglargel}

As $y$ increases 
along the confinement line,
approaching the multicritical point at $y_\text{MCP}$, the ``holes'' in the membranes become larger.
Structurally, these holes are clusters of links on the cubic lattice. 
At the MCP, and on scales much larger than the lattice spacing, they become fractal \textit{loops}\footnote{At the lattice level, the holes (clusters) are not strictly nonintersecting loops, since a site can be connected to more than two occupied links. 
However a visual examination of large loops \cite{somoza2021self} suggests that these self-intersections go away after coarse-graining, so that the infra-red theory is one of nonintersecting loops. 
In the language of loop models, this is equivalent to the statement that the ``four-leg operator'' is renormalization-group irrelevant, or alternatively to the statement that the fractal dimension of self-intersection events is negative \cite{vanderzande1992fractal,jacobsen2003dense}.} with a fractal dimension \cite{somoza2021self}
\be
d_f=1.77(2)
\ee
Slightly below $y_\text{MCP}$, the typical size $\ell(y)$ of the holes is finite (a discussion of scaling forms close to the MCP is given in App.~\ref{app:scalingarg}): 
\ba\label{eq:patchellscaling}
\ell(y) & \sim \f{1}{(y_\text{MCP}-y)^{\nu_S}}, 
&
\nu_S &  \simeq 0.67,
\end{align}
where ${\nu_S=(3-x_S)^{-1}}$ is a critical exponent  of the MCP~\cite{somoza2021self}.
The loops are fractals on scales in between 1 and $\ell(y)$.

Once the holes are patched,
the \textit{patches} inherit some fractal properties from their boundaries.
For example, the typical area of a large patch scales with its linear size $R$ as $R^{1+d_f}$: that is, the effective dimensionality of the patch is greater by one than the effective dimensionality of its boundary, as we might have expected. 

Here, we show that the diverging  lengthscale $\ell(y)$ has implications for  the statistics of the total patched area, $A_\text{patched}$, defined at the end of Sec.~\ref{sec:Isingobservables}.
We compare simulation data with expectations from a scaling argument in that is described in App.~\ref{app:scalingarg}.

Below we describe the approach to the MCP along the confinement line;  qualitatively similar results, but with slightly different exponent values, would obtain if we approached the Higgs phase transition from within the deconfined phase, since $\ell$ also diverges on the Higgs line.

In  Panel~4 of Fig.~\ref{fig:6panels}, the average
fraction of plaquettes that get patched ---
i.e.  $\<A_\text{patched}\>/3L^3$, where  $3L^3$ is the total number of plaquettes in the  lattice --
is  shown as a function of the system size, for several values of $y$. 
The fraction of patched plaquettes  increases slowly as $y\rightarrow y_\text{MCP}$, 
indicating that the typical density and/or size of the holes is growing. 
However
${\lim_{L\rightarrow\infty}  \<A_\text{patched}\>/3L^3}$ 
is finite even at the MCP
(as it must be since by definition this fraction is between zero and one).
At the MCP we may argue 
\be
\f{\<A_\mathrm{patched}(L)\>}{3L^3} = \alpha - {\beta}{L^{-(2-d_f)}} + \ldots.
\ee
where $\alpha$ and $\beta$ are constants.
The first term is dominated by small loops, while the exponent in the subleading term is universal. 
On the confinement line, in the regime ${1\ll \ell(y)\ll L}$, the form  is instead 
\be
\f{\<A_\mathrm{patched}(L)\>}{3L^3} = \alpha - \gamma 
(y_\text{MCP}-y)^{\nu_S(2-d_f)}
 + \ldots,
\ee
where this exponent is $\sim 0.15$ (this is consistent with extrapolated data from Panel 4 Fig.~\ref{fig:6panels} -- data not shown).

Unlike the above average, 
\textit{higher} moments of $A_\text{patched}$ are dominated by the \textit{largest} patches, and do show a critical divergence arising from the divergence of the typical patch size $\ell$. 

When $\ell$ is finite, the normalized cumulants 
$\langle\langle A^k\rangle\rangle/(3L^3)$ for ${k=2,3,\ldots}$
have a finite thermodynamic limit, 
but at the MCP we expect that they are dominated by patches with linear size comparable with $L$, so scale as 
$\langle\langle A^k\rangle\rangle/(3L^3)\sim 
L^{k(1+d_f)}/L^3$.
For $y$ close to $y_\text{MCP}$,
\be
\langle \langle \Apat^k \rangle\rangle
= 
L^{k(1+d_f)} F^{(k)}( L/\ell ),
\ee
where $\ell$ is the lengthscale in (\ref{eq:patchellscaling}) and $F^{(k)}$ is a crossover scaling function.

Panel~5 of Fig.~\ref{fig:6panels}
shows the normalized variance of the patched area,
\be
\langle\langle A_\text{patched}^2\rangle\rangle/(3L^3)
\ee
as a function of the system size, for various $y$.
For $y<y_\text{MCP}$, this quantity has a finite thermodynamic limit.
At the MCP, the data is 
consistent with a power-law divergence in $L$, 
as $\langle\langle A_\text{patched}^2\rangle\rangle/(3L^3) \sim L^{2.2}$.
This exponent is smaller than the expected $L^{2.54}$.
We have not identified the source of the discrepancy. 
Conceivably,  finite size effects associated with the internal structure of a cluster may be large. 
We discuss some aspects of cluster structure in App.~\ref{app:scalingarg}.

Nevertheless, the divergence of the variance of the patched area is a clear sign of a diverging lengthscale for patches at the MCP.

\section{Quasilocality of observables}
\label{sec:locality}

Here we discuss the  locality properties of the line operators $\widetilde{V}_P$ defined in Sec.~\ref{sec:patchingschematic} and of the dual Ising variable.
First we discuss the case where the typical loop (hole) size $\ell$ is finite: in this case there is a notion of locality after coarse-graining, confirming that  patching  successfully defines emergent string operators and the emergent ``duality'' to a local order parameter.
Then in Sec.~\ref{sec:selfdual} we discuss the case where $\ell$ diverges in order to clarify in what sense locality breaks down.

\subsection{Case where $\ell$ is finite}
\label{sec:quasilocality}

Consider the regime where $\ell$ is finite (though perhaps large): this includes the deconfined phase and  the confinement critical regime discussed above (and part of the trivial phase). The following points were discussed more briefly in  Ref.~\cite{somoza2021self}.

How local are the string operators $\widetilde V_P$? 
First let us consider a closed path $P$.
For concreteness, let $P$ be a straight path of length $L$ (on the dual lattice) that winds around a periodic system of size $L$.

When $\ell=0$, the operators (t'Hooft operators) $V_P$
are supported on a single line of plaquettes: i.e. 
the value of 
$V_P(\mathcal{M})$ is a product of $\pm 1$ factors associated with the plaquettes lying on $P$, and so they can be determined by examining the configuration $\mathcal{M}$ only along this line.
For $\ell>0$, the value of 
${\widetilde{V}_P(\mathcal{M})=\pm 1}$ is defined through patching.
In order to determine this value,
we must determine which of the plaquettes on $P$ are occupied in the \text{patched} configuration $\overline{\mathcal{M}}$.

In principle  this can depend on plaquettes of $\mathcal{M}$ that are far from $P$.
However, to determine
${\widetilde{V}_P(\mathcal{M})=\pm 1}$
with a high probability (say $1-\epsilon$) of success,
it is not necessary to patch the entire configuration $\mathcal{M}$.
It is sufficient to patch the loops (holes) in a ``tube'' around $P$. This tube must be  large enough to include, with high probability,\footnote{Probabilities are with respect to the ensemble defining the partition function.} all the loops that are pierced by~$P$.

At first sight we might think that it is sufficient to take a tube of thickness $\mathcal{O}(\ell)$, since this is the typical loop size. However, if the length $L$ of the path is much larger than $\ell$,  then the path will likely encounter some ``rare'' loops of size $\gg \ell$.
This requires us to thicken the tube to a radius that is of order 
\be\label{eq:logscaling}
{R\sim \ell \ln (L/\epsilon)}
\ee
when $L$ is large, in order to have a high probability that all the relevant loops are included in the tube.
(See App.~\ref{app:Vlocality} for more detail.)

In other words, we can accurately approximate $\widetilde V_{P}$ with an operator $\widetilde V_P^\epsilon$ which is a function only of the configuration inside a tube of thickness $R$ around $P$
(and which agrees with $\widetilde V_P$, 
in a random configuration $\mathcal{M}$, 
with a probability very close to 1). 
Note that $R$ is parametrically smaller than the length of the path $P$, so that after sufficient coarse-graining the (t'Hooft) line operator
$\widetilde V_{P}$
can be regarded as local.

Now consider an open path $P$. Letting  $P$ be a straight path between two points $r$ and $r'$,
we can approximate $\widetilde V_P$  by  an operator that depends only on the configuration inside a cigar-shaped region surrounding $P$, whose thickness at the centre is of order ${\ell \ln (|r-r'|/\epsilon)}$.
The thickness of the cigar close to its endpoints, however, can be taken to be only of order 
$\ell$.
In (say) an infinite system,  the operator $\widetilde V_P$, for such an open path $P$,  is dual to ${S_r S_{r'}}$.

The invariance of $\widetilde V_P$ under deformations of the path $P$ means that 
${S_r S_{r'}}$
can be mapped to $\widetilde V_P$ for any path $P$ connecting $r$ to $r'$. 
This, together with the observations above about the cigar-shaped region around $P$, 
gives a sense in which $\ell$ is a typical scale associated with the dual operator $S$. Loosely speaking, a
product such as $S_r S_{r'}$ is robust to  changes in the membrane configuration, so long as these changes are far enough away from $r$, $r'$ and sufficiently local.\footnote{If (i) the region where the membrane configuration is changed is at a distance  $\gg \ell$ from the points $r$, $r'$ and (ii) the complement of the changed region still permits a sufficiently thickened path from $r$ to $r'$ (i.e. the changed region does not ``wrap'' one of the points) then we can choose $P$ so that $\widetilde V^{\epsilon}_P$ is unaffected by the change. In order that $\widetilde V^{\epsilon}_P$ remains a good estimate of $\widetilde V_P$ we may also have to require that the new configuration is not too atypical with respect to the Gibbs measure.}

An alternative way to quantify locality in the dual language would be to reexpress the partition function in terms of $S_r$.\footnote{
Separating $Z$ into topological sectors (see previous sections) we write e.g.   $Z_{+++} = \sum_{\mathcal{M}\in (+++)} e^{-\mathcal{H}[\mathcal{M}]}$
(for the $+++$ sector; other sectors are similar) 
as $Z_{+++} = \f{1}{2} \sum_{\{S_r\}} \sum_{\mathcal{M}\in (+++)} e^{-\mathcal{H}[\mathcal{M}]} 
\chi(S, \mathcal{M})$, where $\chi(S, \mathcal{M})=1$ if $S$ is one of the two spin configurations consistent with the patched configuration $\overline{\mathcal{M}}$ and zero otherwise. Then the effective spin Hamiltonian is given by $e^{-\mathcal{H}'[S]} = 
\sum_{\mathcal{M}} e^{-\mathcal{H}[\mathcal{M}]}
\chi(S, \mathcal{M})$.} Because of patching the effective interactions would be nonlocal, but could be approximated by interactions of finite range $\mathcal{O}(\ell)$ at the cost of incurring only a small error in the free energy density.

We can  also consider the effect on operators $\widetilde V_P$ of changing the patching scheme. If the new scheme retains similar locality properties to the present one,
the change  only dresses $\widetilde V_P$ by quasilocal (on scale $\ell$) functions of the membrane configuration  near $P$'s endpoints. This does not affect the universal asymptotics of correlation functions.

The point above about rare worldline configurations 
(of the gapped anyon), which gave the scaling in Eq.~\ref{eq:logscaling},
should apply generally to the Euclidean path integrals for topological phases. It indicates that generically their emergent line operators must have a thickness that grows logarithmically with their length. This logarithmic scaling is  consistent with
bounds for quantum operators in Ref.~\cite{hastings2005quasiadiabatic}, which considered the definition of dressed line operators, obtained by quasiadiabatic continuation, in discrete gauge theories.

\subsection{At the self-dual MCP}
\label{sec:selfdual}

Right at the self-dual multicritical point, the lengthscale $\ell$ defined by the holes diverges. 
What happens to the observable ${G_{r,r'} \equiv \langle\widetilde V_P\rangle}$ (for a path from $r$ to $r'$)? 

Let us consider the thermodynamic limit  $L\to \infty$ for the system size, with  ${r-r'}$ fixed. 
At first sight we might have guessed that this correlator would fail to have a sensible thermodynamic limit, 
because of contributions from patching loops of size much larger than $|r-r'|$.

In fact, though, we can argue that contributions from loops much larger than $|r-r'|$ are negligible (see App.~\ref{app:scalingarg}), 
so that $\lim_{L\rightarrow\infty}G_{r,r'}$ is well-defined and nonzero even at the MCP. 
This correlator was shown at the MCP in Panel 1 of Fig.~\ref{fig:6panels}
(the average is taken only over non-percolating configurations), and the data suggest  a power law for ${1\ll |r-r'|\ll L}$, with a critical exponent numerically similar to (but presumably distinct from) the Ising exponent.

This power-law scaling of $G_{r,r'}$
 is consistent with scale invariance at the MCP. 
However, at the MCP,
 $\widetilde V_P$ no longer possesses the locality properties that it possessed for finite $\ell$.
First, 
$\widetilde V_P$ can no longer be viewed as a ``string'' operator at the MCP.
This is because, at large $|r-r'|$,
$\widetilde V_P$  is affected non-negligibly by loops of size   $|r-r'|$.
So instead of being a function of the configuration in a  cigar connecting $r$ to $r'$, with a width parametrically smaller than ${|r-r'|}$,
$\widetilde V_P$ is really a function of the configuration in a roughly ball-shaped region whose size is of order ${|r-r'|}$ in all directions.
Similarly, $G_{r,r'}$ is no longer dual to a two-point function in a  spin model with quasilocal interactions.
Finally, at the MCP the exponent of the power law for $G_{r,r'}$  is no longer guaranteed to be independent of the choice of patching scheme.

For these reasons, this ``two-point'' function is no longer such a natural object at the MCP.
Nevertheless it is interesting to see how similar the numerical data for $G_{r,r'}$ at the MCP is to that for $\ell<\infty$. 
This may well be related to the smallness of certain universal amplitudes at the MCP, which we turn to next.

In Table~\ref{tab:sectorsMCP} we show the probabilities for the different sectors, at the MCP. These sectors were defined in Sec.~\ref{sec:vanishingylimit}.
The last column shows the probability of a ``non-patchable'' configuration, i.e.  one with a percolating loop. 
By scale invariance of the ensemble of large loops (holes)  at the MCP \cite{somoza2021self}, this probability is expected to converge to a universal order 1 number, $0< P_\text{perc}<1$, at large $L$.
The data is consistent with this, with
${P_\text{perc}}$ on the order of 0.1. The smallness of this universal number suggests a possible scenario for why  exponents  numerically close to, but distinct from, Landau theory exponents, could arise  at the MCP (see \cite{somoza2021self},~endnote).

\begin{table}[h]
    \centering
    \begin{tabular}{c|ccccc}
        $L$& $P_{+++}$    & $P_{++-}$    & $P_{+--}$    & $P_{---}$   & $P_\mathrm{perc}$ \\ \hline\hline
         8 & 0.219&  0.111&  0.088&  0.077& 0.1061(20)\\
        16 & 0.222&  0.114&  0.089&  0.080& 0.0885(19)\\
        32 & 0.235&  0.115&  0.085&  0.076& 0.0886(19)\\
    \end{tabular}
    \caption{Sector probabilities at the self-dual multicritical point (MCP). Patchable configurations are assigned to a sector $(\sigma_x, \sigma_y, \sigma_z)$ as in Sec.~\ref{sec:vanishingylimit}. These sectors have probability $P_{\sigma_x, \sigma_y, \sigma_z}$. By symmetry, $P_{++-}=P_{-++}$ etc.
    Percolating (non-patchable) configurations occur with nonzero universal probability $P_\text{perc}$ at large $L$.
 (By contrast, ${\lim_{L\to\infty} P_\text{perc}=0}$  in the region of the phase diagram where $\ell$ is finite.)
    }
    \label{tab:sectorsMCP}
\end{table}

\section{Extensions of the patching idea}
\label{sec:extensions}

The approach of this paper can be extended to numerous other models that have a representation in terms of ``closed'' geometric objects.
It can be applied either as a way of diagnosing a topological phase or as a way of probing phase transitions. 
The geometric objects need not necessarily be two-dimensional membranes.
Here we list some possible examples for future investigation.

\subsection{Three dimensions: gauge theories and amphiphilic membranes}

The closest to the present setting are other 3D models that have a sign-free representation in terms of membranes representing, say, worldsurfaces of ``electric'' flux lines.
Discrete gauge theories with other gauge groups give  modified kinds of membranes. 
For example, $\mathbb{Z}_3$ gauge theory gives membranes that carry an orientation degree of freedom.  Allowing multiple species of matter field corresponds to giving the $e$ anyon, and therefore the ``holes'' in the membranes, an additional label.

It should be emphasized, however, that the patching approach is more general than the gauge theory context. 
For example, it could be applied  to configurations of amphiphilic membranes \cite{huse1988phase,cates1988random,roux1990light,roux1992sponge,roux1995sponge,pelitiamphiphilic,huse1991sponge,roux1992sponge,lipowsky1991conformation} in order to detect the phase transitions out of  the ``symmetric sponge phase'' (deconfined phase). 

In principle such models can be in the spatial continuum, rather than on a lattice.
It would be interesting  to explore how far one can depart from standard gauge theory Hamiltonians while maintaining the phase diagram topology of Fig.~\ref{Fig:phasediagram}.

\subsection{3+1 dimensions}

In a 3+1 dimensional discrete gauge theory with gapped matter, we can again patch the 2D worldsurfaces of electric flux lines. This will allow us to detect the emergent 1-form electric symmetry \cite{gaiotto2015generalized} of the deconfined phase.

It is also interesting to consider cases 
(relevant for example to 3+1D Higgs transitions) 
where the ``membranes'' that we patch are three-dimensional, or more generally a $d+1$-dimensional theory where we patch $d$-dimensional surfaces. 
The patched surfaces can then be interpreted as domain walls for a dual Landau-like field $\varphi$.
 In 3+1 dimensions we can have a non-fine-tuned  phase transition (with no symmetries in the UV) where  this dual Landau field is free in the IR.

\subsection{Repairing loops}

\begin{figure}
    \centering
    \includegraphics[width=0.55\linewidth]{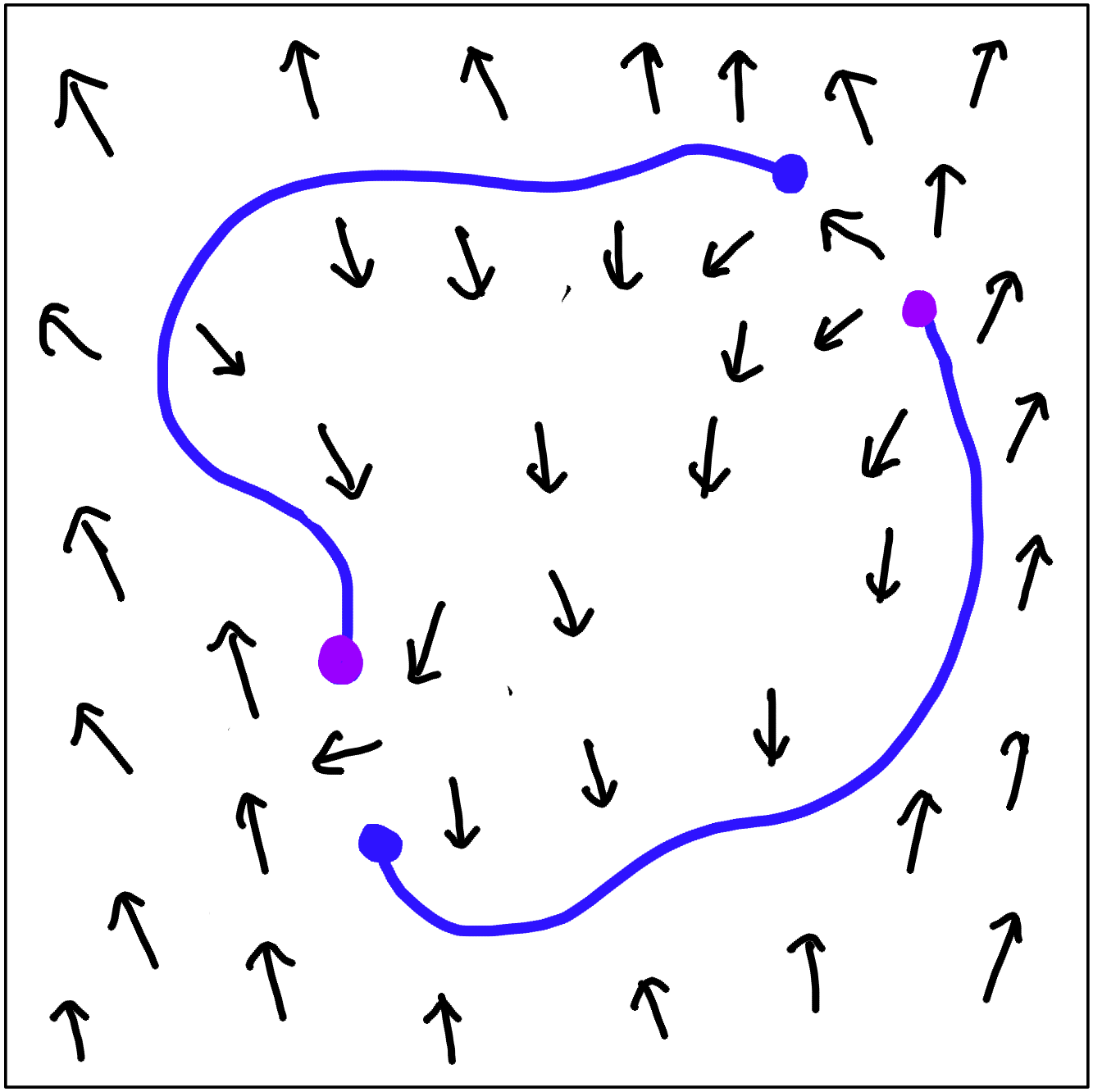}
    \caption{ A configuration of XY spins in a modified XY model, with 1/2-vortices (dots) that are connected by strings (blue) where the phase (indicated by arrows) jumps by $\simeq \pi$. A repair algorithm for the strings allows additional correlators to be defined.}
    \label{Fig:halfvortices}
\end{figure}

It is interesting to consider repairing \textit{one}-dimensional lines, for example in 2D. 

Refs.~\cite{korshunov1985possible,lee1985strings,shi2011boson,serna2017deconfinement} analysed a modified Hamiltonian for the 2D classical XY model, in which the interaction energy for a pair of adjacent spins $i$ and $j$ has two minima, one at relative angle ${\theta_i-\theta_j=0}$, and one at ${\theta_i-\theta_j=\pi}$:
\be
\beta H = - \sum_{\<ij\>}  
\left[
\Delta \cos\lf \theta_j-\theta_j \ri
+ 
J\cos \lf 2 (\theta_j - \theta_i)\ri
\right].
\ee
When $J$ and $\Delta$ are both large, this model has quasi-long-range-order for $e^{i\theta}$ (the superfluid phase). 
If $\Delta$ is now decreased (still at large $J$), 
the energy cost of certain string excitations, across which $\theta$ jumps by $\pi$, is decreased --- see Fig.~\ref{Fig:halfvortices}.
In the limit of infinite $J$ these strings are closed, 
while at finite $J$ they can terminate at half-vortices (Fig.~\ref{Fig:halfvortices}).
Varying $\Delta$ at fixed large $J$, we may cross a phase transition at which the strings proliferate. 
This is a transition from the superfluid into a pair superfluid, where $e^{i\theta}$ is disordered (thanks to the strings, across which $e^{i\theta}$ changes sign) but $e^{2i\theta}$ has quasi-long-range order \cite{shi2011boson}.  

The simplest picture for this transition is to neglect the half-vortices, since at large $J$ they appear only in bound pairs. 
Then the strings are treated as closed loops, and are analogous to Ising domain walls. 
As a result, the phase transition
at large $J$ (between superfluid and pair superfluid) has Ising exponents  \cite{korshunov1985possible,lee1985strings}. There are also interesting phase transitions at smaller $J$~\cite{shi2011boson,serna2017deconfinement}. 

This and similar 2D models could  be studied from the ``patching'' point of view of the present paper, 
explicitly repairing the loops with an algorithm that pairs (matches) nearby half-vortices, and adds a corresponding segment to the string. 
Assuming that this repair process is sufficiently local, it would for example allow the Ising anomalous dimension $\eta=1/4$ to be measured directly using a (nonlocal) correlation function, despite the fact that there is no local operator with this anomalous dimension in this theory.

Note that the emergence of closed (unoriented) loops here implies the emergence of a $\mathbb{Z}_2$ symmetry. If one direction is viewed as time, this is the conservation of the number of loop segments, modulo~2. 

It would be interesting, both in the above case and in the standard XY model/Coulomb gas, to study the relation between the thermodynamic binding/unbinding of vortices, and 
the success or failure of a given algorithm in finding a quasi-local pairing of vortices and antivortices.

\subsection{Application to experiments in quantum devices}

So far we have discussed $\mathbb{Z}_2$ gauge theory in the language of 3D classical statistical mechanics. 
But this gauge theory also describes the quantum statistical mechanics of the simplest $\mathbb{Z}_2$ spin liquids/topologically ordered states  in two dimensions, including the toric code \cite{read1989statistics, kivelson1989statistics,
read1991large, wen1991mean, senthil2000z, moessner2001short,kitaev2003fault}. 

Recently it was possible to realize the toric code ground state experimentally \cite{satzinger2021realizing,bluvstein2022quantum}.
A key question is how to efficiently detect deconfinement in experiments of this type by using measurements of the spins. 
One approach is the  Fredenhagen-Marcu order parameter, which involves a ratio of correlation functions of ``bare'' Wilson line operators \cite{fredenhagen1986confinement, huse1991sponge,gregor2011diagnosing,xu2024critical}. However this involves measuring exponentially small quantities (because of nonuniversal short-distance effects close to the Wilson line), implying a signal-to-noise problem in the large size limit. 
Therefore we suggest that an effective method may be instead to directly define \textit{emergent} string correlators by applying the configuration ``repair'' idea of Ref.~\cite{somoza2021self} to 2D quantum measurements, rather than 3D Monte Carlo data. Below we give  a schematic indication. 

After this  work was completed, we learned of  a related proposal for detecting topological order from 2D quantum measurement snapshots  in Ref.~\cite{cong2022enhancing} (where an RG-inspired procedure for annihilating anyons is used, rather than minimal-weight matching).

A complete measurement of all the qubits making up the quantum state, say in the $\sigma_x$ basis, gives a 2D configuration $\{\sigma_x\}$. 
(This configuration is of course random, according to Born's rule, even though we assume that the initial wavefunction is prepared deterministically.)
This configuration is equivalent to a single 2D \textit{slice} through the 3D configurations discussed above.
Since we only have 2D information, the problem of ``repairing'' the configurations is  different to that discussed above, although the general idea is similar.

For idealised wavefunctions (e.g. the toric code), such a measurement gives a $\sigma_x$ configuration that can be viewed as a collection of closed loops (i.e. electric flux lines) in the plane. 
In this setting, the value of a magnetic string operator $V_P(\{\sigma_x\})$ can be defined in a standard way for a given $\sigma_x$ configuration.
The definition is the planar version of that in Sec.~\ref{sec:vanishingylimit}: each electric flux line that is crossed by $P$ contributes a minus sign to $V_P(\{\sigma_x\})$.
The average of this operator for long path $P$ be used to distinguish the deconfined from the confined phase \cite{kitaev2003fault} [point \textit{(ii)} in Sec.~\ref{sec:diagnosingdeconf}].  
Here, averaging requires repeatedly measuring independent instances of the wavefunction.

For a more generic wavefunction the $\sigma_x$ configurations obtained by measurement will not form closed loops. However, if we are sufficiently close to the idealized limit, we can  define emergent string operator measurements via 
\be
\widetilde V_P(\{\sigma_x\})\equiv V_P(\{\overline{\sigma}_x\}),
\ee
where $\{\overline{\sigma}_x\}$ is a repaired configuration in which dangling-end defects are eliminated. 
The annihilation of pairs of defects is a standard idea in error correction \cite{dennis2002topological}: in  the most basic standard algorithm, $\overline{\sigma}_x$ is obtained from $\sigma_x$ by flipping a minimal set of spins that removes all dangling ends.
This geometrical algorithm is used here for a different purpose, namely to define  functions $\widetilde V_P(\{\sigma_x\})$ for any configuration.

At first glance, having defined 
$\widetilde V_P(\{\sigma_x\})$, we can now use expectation values of 
$\widetilde V_P(\{\sigma_x\})$ 
(averaged over complete measurements of many instances of the state)
to demonstrate deconfinement in the same way as in the idealized case. 

However, it is necessary to first verify that the repair process gives a sufficiently local operator ${\widetilde V_P(\{\sigma_x\})}$. 
In the \textit{three}-dimensional case, we were able to analyze the locality properties of the patching using percolation concepts and to confirm that patching succeeds all the way up to the Higgs phase boundary. 
We leave it for the future to analyse the locality of this \textit{two}-dimensional repair process.

\section{Outlook \& comments on the MCP}
\label{sec:outlook}

This paper advocates patching as constructive way of understanding emergent higher symmetries and as a useful tool in simulations. (For example, dimensionless ``Binder cumulants'' of the dual order parameter may be used to accurately locate the transition.)

We have focused on the case of a simple lattice gauge theory, but patching and its generalizations could be used to construct emergent string operators and dual order parameters in much more general models. It would be interesting to make further numerical explorations.
Sec.~\ref{sec:extensions} has discussed some examples including   models of membranes in the spatial continuum, a generalized XY model, and quantum simulations of topological states.

A key question in any patching scheme is the locality of the patching process. 
In the patching algorithm that we use, each ``hole'' (loop) is patched separately. 
As a result, the locality properties of the algorithm are analyzable using standard ideas from geometrical phase transitions, showing that nontrivial emergent line operators can be  constructed in the entirety of the part of the phase diagram where they are expected to exist in the~IR.

This loop-by-loop patching scheme is different from a ``minimal weight'' algorithm, which would find the minimal-area set of patches consistent with a given set of loops. 
One issue with the  latter algorithm is that it involves an optimization problem that may be highly nonlocal even in a configuration where all loops are finite. 
The locality properties of such  a patching procedure would therefore require further analysis.

 Gauge fixing in the minimal gauge \cite{fradkin1979phase} 
is closely related to minimal-weight patching (in the appropriate membrane ensemble). 
In a finite system it differs in the way global excitations are treated.
Let us comment on this briefly.
The relevant membrane ensemble for gauge fixing is the magnetic ensemble, which is dual to the one we have primarily discussed.
Loosely speaking, the relation is that, after gauge fixing, 
the domain walls of the gauge-fixed Higgs field configuration $\tau^{\text{gauge-fixed}}_i$
(cf. Eq.~\ref{eq:Z3Dgaugerep}) define a configuration $\mathcal{M}_1$ of closed membranes. 
In an infinite system, these are precisely the membranes that are obtained by ``patching'' an initial configuration of (possibly open) membranes defined by the values of ${\sigma_{ij}\tau_i\tau_j}$, using a minimal-weight algorithm.
In the finite system, however, 
the configuration $\mathcal{M}_1$ can differ from the (minimal-weight) patched configuration $\overline{\mathcal{M}}$ by system-spanning membranes. 
As a result, the gauge fixing procedure does not in general give access  to the operators $\widetilde V_P$ in the finite periodic system.
 However, in the infinite-system limit, minimal-gauge-fixing corresponds to a certain choice of patching scheme.

An alternative tool for diagnosing deconfinement is the Fredenhagen-Marcu order parameter
\cite{fredenhagen1986confinement,gregor2011diagnosing}, whose critical behavior was recently studied numerically \cite{xu2024critical}.
This observable uses a ``bare'' string operator, rather than a dressed one, and as a result generally suffers an exponential signal-to-noise problem in Monte Carlo. However the  scaling of the FM order parameter could be understood via an extension of the arguments in Sec.~\ref{sec:locality} and the appendices, by writing the bare string operator as the product of the dressed operator and a contribution from patches.

In this paper we have considered the gauge theory with periodic boundary conditions, but it would also be interesting to explore boundary phenomena. 
Recently it has been argued that Higgs phases can often be thought of as symmetry-protected topological states protected by both conventional and higher-form symmetries \cite{verresen2022higgs}, with associated boundary states.
(See Refs.~\cite{greensite2022symmetry,nussinov2005derivation} for other approaches to distinguishing  Higgs and confining regimes.)

Inside the deconfined phase, 
there are two emergent one-form symmetries, related by e--m duality.
One of these symmetries survives at the Higgs transition, and the dual symmetry survives at the confinement transition. 
The approach in the present paper only gives access to one set of string operators at a time, because it is necessary to pick a basis (electric or magnetic) in order to formulate the partition function as a membrane ensemble. 
The phase diagram of the standard $\mathbb{Z}_2$ gauge theory has an exact self-duality, so it is straightforward to change between the two  bases. 
For more general models (e.g. models in the spatial continuum) the analog of this basis change may be more nontrivial.

\subsection{Comments on self-dual multicritical point}

Finally let us discuss the multicritical point in Fig.~\ref{Fig:phasediagram}. In addition to the confinement and Higgs transition lines, 
the phase diagram of the gauge theory contains the self-dual multicritical point (MCP) where they meet.
The MCP is not yet understood as a conformal field theory, and is an interesting target for further study.

Since the MCP is the meeting point of two critical lines with Ising exponents, at first sight a natural guess would be that it has XY exponents \cite{huse1991sponge}. 
More recently, this has been argued by Ref.~\cite{bonati2022multicritical}, 
on the basis of the numerical similarity of exponents observed in Ref.~\cite{somoza2021self}. 
However, challenges to this theoretical interpretation were discussed in Refs.~\cite{somoza2021self,oppenheim2023machine}.
To begin with, the fixed point associated with the MCP cannot be the fixed point of the XY model
(for example because the adjacent phases do not match).
In addition,  the MCP cannot be an orbifold\footnote{Here orbifolding refers to gauging with a flat gauge field (which leaves correlators of gauge-invariant operators unchanged and eliminates non-gauge-invariant operators from the spectrum). The term is used in a different sense in some other contexts \cite{schmaltz1999duality}.} of the XY fixed point (i.e. an ``XY$^*$ transition'') as this would not give the correct universal properties of the topologically ordered phase immediately adjacent to the MCP \cite{somoza2021self}.

Ref.~\cite{oppenheim2023machine} has directly demonstrated (using simulations) that there is no U(1) current operator at the MCP CFT, showing that the MCP does not have an emergent global U(1) symmetry.
By contrast, the usual ``XY$^*$'' transition has a U(1) symmetry.

These observations do not rigorously exclude the possibility that the exponents of the MCP could be equal to exponents from XY. 
But (as observed in \cite{somoza2021self}) if this was the case, it would have to be due to an entirely new type of relationship between conformal field theories, and at present we are not aware of any  proposed mechanism. 
Therefore at present the simplest hypothesis is that the exponents of the MCP are simply numerically close to those of the XY model.

In this paper we have focussed mainly on the part of the phase diagram 
where the ``hole size'' $\ell(x,y)$ is finite, so that
patching allows a duality to a model with a local order parameter.
This works for the confinement transition (as well as in the deconfined phase).
The equivalent process works for the Higgs transition, in the dual representation.
But at the multicritical point (where Higgs and confinement lines meet), 
$\ell$ diverges. As a result, patching does not give a duality to a model with a local order parameter at the MCP. This is another reason why, a priori, we do not expect a Landau-like description of the MCP.

Instead, any continuum description needs to take account of the braiding statistics in the adjacent deconfined phase. 
The simplest framework that can account for the anyonic braiding statistics of $\mathbb{Z}_2$ topological order using continuum (as opposed to lattice) field theory is a mutual Chern-Simons (or ``BF'') theory, with two U(1) gauge fields, 
$a^1$ and $a^2$, with a Chern-Simons term 
${i\f{k}{2\pi}\epsilon_{\mu\nu\lambda} a_\mu^1 \partial_\nu^{\phantom{1}} a_\lambda^2}$ at ${k=2}$.
These gauge fields are coupled
to two matter fields,
whose quanta represent e and m anyons respectively \cite{hansson2004superconductors,diamantini2008topological,kou2008mutual,mcgreevy2017tasi}:
\ba\notag
\mathcal{L}_{\rm CS} = & \,
  \f{i}{\pi} \epsilon_{\mu\nu \lambda} a^1_\mu \partial_\nu a^2_\lambda +  \f{1}{2} \sum_{I = 1,2}  \hspace{-0.5mm}
 |(\partial - i a^I) z^I|^2  
 \\ + & \f{m^2}{2}  \hspace{-0.8mm} \sum_{I = 1,2}
 \hspace{-0.5mm} | z^I|^2
  +  
 \lambda \hspace{-0.5mm} \lf |z^1|^4 + |z^2|^4 \ri 
 + \lambda'   |z^1|^2 |z^2|^2
 + \ldots\notag
 \end{align}
However, if we want a putative description of the MCP, then we must also add monopole operators $\mathcal{M}^{(I)}$, for each of the gauge fields $I=1,2$, to this continuum Lagrangian.
This is necessary to avoid an unphysical enlargement of symmetry, or equivalently to remove an unwanted degeneracy in the spectrum of anyons:
\be
\mathcal{L} = \mathcal{L}_{\rm CS} + 
\kappa \sum_{I=1,2} \lf\mathcal{M}^{(I)}+\mathcal{M}^{(I)*}\ri. 
\ee
This issue was commented on in Ref.~\cite{somoza2021self}.
In App.~\ref{app:fieldtheoryissues} we review the Chern-Simons approach in more detail in order to clarify why the multicritical point is challenging to describe using continuum field theory.
(See a very recent preprint for a discussion of $\mathbb{Z}_k$ for larger values of $k$ \cite{shi2024analytic}.)
App.~\ref{app:fieldtheoryissues} also discusses a (speculative) alternative way to remove the unwanted anyon degeneracy, using a replica-like limit in the Chern-Simons theory.

\acknowledgments 
 We thank 
Snir Gazit, 
Lior Oppenheim, 
Zohar Ringel, 
Balt van Rees,
Slava Rychkov,
Chong Wang, and
Dominic Williamson
for discussions. 
A.S. acknowledges support by MCIN/AEI/10.13039/501100011033 through project  PID2022-139191OB-C31 “ERDF A way of making Europe".

\appendix

\section{Scaling arguments}
\label{app:scalingarg}

\subsection{Lengthscales close to the self-dual MCP}

The multicritical point lies on a self-dual line  in the phase diagram.
Choosing the phase diagram coordinates $(y, y')$ where 
\be
y' \equiv
\f{1-x}{1+x},
\ee
self duality acts as 
\be
y \longleftrightarrow y'.
\ee
Self-duality (which can be treated as a global $\mathbb{Z}_2$ symmetry in the infra-red) can be used to classify operators at the MCP. 
There is an RG relevant, self-duality-preserving  operator denoted $S$, with scaling dimension $x_S\sim 1.5$, and an RG relevant self-duality-odd (anti-self-dual) operator denoted $A$, with scaling dimension $x_A\sim 1.22$ \cite{somoza2021self}.
Writing ${\delta y = y-y_\text{MCP}}$,
the corresponding perturbations of the MCP are ${\lambda_+ \equiv \delta y+\delta  y'}$
and ${\lambda_- \equiv \delta y'-\delta y}$, respectively, and their RG eigenvalues are ${ 3-x_{S,A}}$.
We let ${\nu_{S,A} \equiv (3-x_{S,A})^{-1}}$.

Turning on a  $\lambda_+$ perturbation with $\lambda_+<0$ leads to the deconfined phase, while turning on 
$\lambda_-$ leads to either the confined or the Higgs state, depending on $\operatorname{sign} \lambda_-$. By standard RG results,
the shape of the confinement and Higgs phase transition lines, close to the MCP, are given by 
\be\label{eq:transitionlineshape}
\lambda_- \sim  \pm |\lambda_+|^{\nu_S/\nu_A},
\ee
so roughly 
\be
\lambda_- \sim \pm |\lambda_+|^{1.2}.
\ee
Therefore the confinement and Higgs phase transition lines are asymptotically parallel as they approach the MCP, but this cusp behaviour is rather weak since the above exponent is not much larger than 1.

Note that a crude way to obtain the above is to define lengthscales 
\be\label{eq:lengthscalesnrMCP}
\xi_\pm \equiv  |\lambda_\pm|^{-\nu_{S,A}},
\ee
which are the scales at which a given perturbation would renormalize to an order 1 value (in the absence of the other perturbation). The phase transition lines occur where $\xi_+$ and $\xi_-$ are of the same order (otherwise one of the two perturbations dominates, and we end up in the interior of one of the phases). 

Therefore at a point \textit{on the confinement phase transition line}, but close to the MCP, we have a large lengthscale 
${\ell\sim \xi_+\sim \xi_-}$.
Since, on the confinement line, we have (from Eq.~\ref{eq:transitionlineshape})
\be
\lambda_+ = 2\, \delta y + \mathcal{O}(\lambda_+^{\nu_S/\nu_A}),
\ee
this lengthscale is
\be\label{eq:appcrossoverscale}
\ell(y) \sim (y_\text{MCP}- y)^{-\nu_S}.
\ee
This is the characteristic lengthscale for the crossover from the universality class of the MCP to universality class of the 
Ising$^*$ confinement transition. 
(Analogous statements apply for the crossover on the Higgs line.)
$\ell(y)$ is also the lengthscale associated with patches, i.e. the scale beyond which larger patches become exponentially rare. 

\subsection{Fractal structure of patches}

When $\ell\gg 1$ we have large patches.  We now give scaling arguments using ideas from geometrical critical phenomena (see e.g. Ref.~\cite{stauffer2018introduction}) for (i)  the typical area of a large patch and (ii) the moments of the total area of all patches. 

First, consider a single large patch whose boundary ``loop'' (link cluster) has linear size $l$.
The total length of this loop --- total number of ``occupied links'' making  up the cluster --- scales as $l^{d_f}$ with 
\be
d_f = 1.77(2).
\ee
Therefore within a ball of volume $l^3$ centred on the centre of mass (CM) of the loop, the average density of occupied links is $\sim l^{d_f-3}$. 
Approximating the distribution of occupied links within this volume as uniform, the number of occupied links within a spherical shell of thickness $\dd r$ at radius $r$ scales as $N(r)\dd r = l^{d_f-3} r^2 \dd r$.

Recall that, to define the patch (Sec.~\ref{sec:patchingalgo}), we first associate a triangle with each occupied link. This triangle connects the link to the loop's centre of mass. Any link of the \textit{dual} lattice that pierces an odd number of triangles then gives rise to one of the plaquettes in the patch.

Let $N_\text{dual}(r')\dd r'$ be  the number of dual links that lie within the spherical shell with radii ${[r', r'+\dd r']}$ and which pierce an odd number of triangles.
The density of patched plaquettes is then ${\rho_\text{patch}(r') = N_\text{dual}(r')/r'^2}$.

To compute $N_\text{dual}(r')$, we add up contributions from different triangles, associated with occupied links at various possible radii $r$ (where $r'< r \lesssim l$).
To begin with we neglect double-counting, i.e. the possibility of a dual link piercing more than  one triangle.
This will be valid in a range $r'\gg r_c$ where
$\rho_\text{patch}(r') \ll 1$.
The scale $r_c$ defines a ``dense core'', within which 
different triangles ``overlap'', double-counting cannot be neglected, and 
$\rho_\text{patch}(r')$ is of order 1. However, $r_c\ll l$, so the dense core has a subleading effect on the total area of the patch.

Consider a triangle associated with an occupied link at radius $r$. 
Since the triangle gets thinner towards the CM, the part of it that lies within the shell ${[r', r'+\dd r']}$  (with $r'<r$) has area $\sim (r'/r) \dd r'$,
and on average  is pierced by order $(r'/r) \dd r'$ dual links within the shell ${[r', r'+\dd r']}$. 
That is, a triangle whose outer edge is at radius $r$ typically makes a contribution $\sim (r'/r) \dd r'$ to $N_\text{dual}(r')\dd r'$.

Adding up such contributions,
\ba
N_\text{dual}(r') & \sim  \int_{r'}^l \dd r N(r) (r'/r)\\
& = l^{d_f-3} r' \int_{r'}^l \dd r\, r  \\
& \sim l^{d_f-1} r' H(r'/l), \label{eq:Ndualfinal}
\end{align}
where $H(x)\rightarrow 1$ for $x\ll 1$ and $H(x)\rightarrow 0$ for ${x\gg 1}$. 
(In our caricature, where the density is uniform, ${H(x) = [1-x^2]}$, but this is not accurate.)
This gives the density
\be\label{eq:rhopatch}
\rho_\text{patch}(r') = {l^{d_f-1}} H(r'/l)/r'.
\ee
So far we have neglected double counting. 
From (\ref{eq:rhopatch}) we see that the radius of the dense core, within which $\rho_\text{patch}$ is cut off at an order 1 value,
is
\be
r_c \sim l^{d_f-1} \sim l^{0.77(2)}.
\ee
Since this exponent is smaller than one, the size of the dense core is much smaller than the total size of the patch, $r_c \ll l$. Integrating the density, we find that the typical area of a patch of linear size $l$ is
\ba\label{eq:patchsizescaling}
A(l) &\sim l^{d_f+1} \lf 1 - \mathcal{O}(l^{-2(2-d_f)}) \ri \\
&\sim l^{2.77} \lf 1- \mathcal{O}( l^{-0.46}) \ri,
\end{align}
where the subleading correction is due to the dense core. 
(In principle there will be other sources of subleading corrections, e.g. due to irrelevant corrections to scaling, that we have not considered.)

Next, consider the total patched area in a membrane configuration.
We begin with the MCP, where there are clusters on all scales up to the system size $L$.

For a crude picture, think of $A_\text{patched}$ as a sum over contributions from clusters on a range of lengthscales $l_\tau=2^\tau$ for ${\tau = 1, \ldots, \log_2 L}$:
\be\label{eq:Apatchedscales}
\Apat = \sum_{\tau=1}^{\tau_\text{max}} \sum_{i=1}^{n_\tau} A^{(\tau)}_i.
\ee
Here $i$ indexes the clusters at a given scale and $n_\tau$ is the number of clusters at a given scale. A caricature that captures the scaling is imagine dividing up the system into $l_\tau$-sized blocks, and to think of $n_\tau$ as a sum of roughly independent contributions, each of size $\mathcal{O}(1)$, from each block
(in reality these contributions are not independent, but that will not matter for the following). We have 
\be
n_\tau \sim (L/l_\tau)^3.
\ee

Using (\ref{eq:patchsizescaling}) above for the typical size $A(l)$ of a patch, we have for the average
\ba
\f{\< A_\text{patched} \>}{3L^3} \sim & 
(3L^3)^{-1} \sum_{\tau=1}^{\log_2 L} 
(L/l_\tau)^3
A(l_\tau) 
\\
\sim & \alpha - \beta L^{-(2-d_f)}
\end{align}
as stated in the main text. (The  $\alpha$ term
is dominated by UV contributions.)
If  $\ell$ is finite and ${1\ll \ell \ll L}$ then the sum will instead be cutoff at ${l_\tau\sim \ell}$, so that the subleading term is  of order
${\ell^{-(2-d_f)}=(y_\text{MCP}-y)^{\nu_S(2-d_f)}}$:
\be
\lim_{L\rightarrow\infty}  \f{\<A_\text{patched}\>}{3L^3} 
= \alpha - \gamma (y_\text{MCP}-y)^{0.15} + \ldots 
\ee

Writing higher moments of
$A_\text{patched}$  as a similar sum over scales suggests that they are instead dominated by the largest patches, which are of scale $L$. There are $\mathcal{O}(1)$ patches of this scale with area and fluctuations of order $A(L)$, so that 
\be
\langle \langle A_\text{patched}^k \rangle\rangle \sim A(L)^k 
\sim L^{k(d_f+1)}.
\ee
Moving along the confinement line away from the MCP slightly, a standard rescaling argument gives a scaling form in terms of the crossover scale $\ell$ in Eq.~\ref{eq:appcrossoverscale}:
\be
\langle \langle A_\text{patched}^k \rangle\rangle 
\sim L^{k(d_f+1)} F^{(k)}(L/\ell).
\ee

The above carries over straightforwardly to the scaling close to the Higgs line (rather than the MCP) where $\ell$ also diverges, but with the usual Ising correlation length exponent $\nu_\text{Ising}$, and where the loops have a distinct fractal dimension (that of Ising worldlines, $d_f'= 1.7349(65)$ \cite{winter2008geometric}).
We can also consider a more general point in the vicinity of the MCP, in which case the scaling functions depend not just on $L/\ell$ but on $L/\xi_+$ and $L/\xi_-$ (Eq.~\ref{eq:lengthscalesnrMCP}) separately.

\subsection{Note on locality of $\widetilde{V}_P$}
\label{app:Vlocality}

We give slightly more detail on the discussion of the local truncation of $\widetilde V_P$ in Sec.~\ref{sec:quasilocality}.

Consider the closed path case. The exponential suppression of loops of size $R\gg \ell(x,y)$
means that if we thicken the tube to a scale $R$ (with ${\ell\ll R\ll L}$),
the probability that there is a loop pierced by $P$ which we fail to include in the tube is less than ${A L e^{-R/\ell'(x,y)}}$
(where $\ell'(x,y)$ is proportional to  $\ell(x,y)$ when they are both large, and $A$ is a constant). 
Therefore in order to have a probability of error that is less than $\epsilon$, it is sufficient to take
\be 
R > \ell'(x,y) \, \ln \lf \f{A L}{\epsilon}\ri.
\ee
This means that we can approximate $\widetilde{V}_P$
with an operator $\widetilde{V}^\epsilon_P$
that is a function only of the plaquettes inside a tube of radius $R$.
These operators agree in a configuration with high probability, so their correlation functions are also approximately equal.

\section{Higgs transition as a percolation transition: replica approach}
\label{sec:percolationanalytic}

In this Appendix, following on from the comments in Sec.~\ref{sec:diagnosingdeconf},
 we  discuss the relation between 
the Higgs transition line and the hole percolation transition line (shown in Fig.~\ref{Fig:phasediagramwithpercolation}) where the typical size of holes diverges.
Our aim is to connect the percolation transition to the standard ``replica'' formalism used for other geometrical critical phenomena \cite{cardy1996scaling,nahum2012universal}.
(For simplicity we discuss only values of $x$ that lie above the multicritical point.)
This replica formulation does not rigorously prove anything, but it gives some intuition for why the percolation transition and the Higgs transition can coincide without fine-tuning.
In other words, for why the 
phase diagram topology in 
Fig.~\ref{Fig:phasediagramwithpercolation} --- with the percolation line lying on top of the Higgs line --- is stable to small perturbations of the model or of the protocol for defining clusters.

At the end of this section we also briefly describe, without using replicas, the renormalization group property of the Higgs transition that ensures this topology is stable.

Recall that the percolation transition is defined by the divergence of the typical hole size, where a ``hole'' is defined as a connected cluster of occupied links
(Sec.~\ref{sec:patchingalgo}).\footnote{In simulations the  percolation transition may be detected via the appearance of system-spanning clusters
\cite{somoza2021self}.}
Therefore diagnosing percolation requires access to nonlocal ``geometrical'' correlation functions. These cannot be written down as local correlators of the orginal degrees of freedom, but they can be written down using the replica trick 
\cite{cardy1996scaling}.
We follow  an analogous discussion for vortices in the XY model \cite{nahum2012universal}.

Consider the lattice action for the gauge theory (\ref{eq:Z3Dgaugerep}) written schematically as
\be
\mathcal{S} = -  K \sum_\square \prod \sigma  - J \sum_{\<ij\>} \sigma_{ij} \tau_i \tau_j.
\ee
In order to access geometrical correlators for clusters, we extend this action to a hierarchy of actions indexed by a natural number $N$: 
ultimately we are interested in correlation functions in the ``replica limit'' ${N\to 1}$. We replace the field $\tau_i$ with an $N$ component vector $\vec \tau_i$, with components $\tau_i^a$ for $a=1,\ldots, N$:
\be\label{eq:fieldtheoryreplica}
\mathcal{S} = -  K \sum_\square \prod \sigma  - J \sum_{\<ij\>} \sigma_{ij}  \vec \tau_i.\vec \tau_j.
\ee
We sum over values of ${\vec \tau}_i$ in which a single component is nonzero, and is equal to $\pm 1$: i.e. ${\vec \tau = (\pm 1, 0, 0 \ldots, 0)}$, ${\vec \tau = (0, \pm 1, 0, \ldots, 0)}$, etc.  
This replicated theory now has a nontrivial \textit{global} symmetry, namely $(S_N\ltimes \mathbb{Z}_2^N)/\mathbb{Z}_2$.

The original partition function is reproduced in the replica limit $N\to 1$. 
However, the analytic continuation from larger values of $N$ gives access to geometrical observables that are nonlocal in the original formulation.
For example, the probability of two sites lying on the same connected cluster of links is given by 
\be\label{eq:pconn}
P(\text{$i$ connected to $j$}) 
= \lim_{N\to 1} \, (N-1)^{-1} \< Q^1_i Q^1_j\>,
\ee
where we have defined the lattice operator 
\be\label{eq:appQfield}
Q^a_j = (\tau_j^a)^2 - 1/N.
\ee
This may be shown by examining the high temperature (``strong coupling'') expansion, in which each  cluster of connected links  in the expansion acquires a color index that runs over $N$ values. This is similar to other constructions 
\cite{cardy1996scaling, nahum2012universal} so we omit details.

A percolating phase is therefore one where the two point function in Eq.~\ref{eq:pconn} is long-range ordered, which we write as ${\<Q\> \neq 0}$. In other words,  the $S_N$  permutational symmetry of the replica theory is broken in the percolating phase.

If we start at small $y$ in the deconfined phase,   $\vec \tau$ is uncondensed, and $\<Q\> = 0$.
Now, a priori, we could achieve percolation in two different ways.

First is the way that in fact happens in this model: as we increase $y$, we pass through the Higgs transition, at which  $\vec\tau$ condenses.
Since $\vec \tau$ transforms nontrivially under $S_N$, the natural expectation is that the condensation of $\vec \tau$ induces ``subsidiary'' long range order in the composite  field $Q$ (i.e. percolation).
That is, the Higgs and hole-percolation transitions coincide.

In principle, however, we can imagine another phase.  This is a ``pair condensate'', in which $Q$ is condensed ($\<Q\>\neq 0$), but in which $\vec \tau$ is not condensed.
This  corresponds to a situation in which clusters percolate \textit{within} the deconfined phase.
Though this does not  happen in the present model, for the present definition of clusters,
it is easy to see that such a phase can be achieved if  we modify the definition of clusters sufficiently.

For example, we could modify the definition of clusters in the following way. In turn, this defines a modified patching scheme.

Let $\mathcal{O}$ denote the set of occupied links. 
(Recall that in the definition above, a cluster is a connected subset of $\mathcal{O}$.)
Choose some probability
 $p  \in [0,1]$,
 and form another  (random) set of links,  $\mathcal{O}'$, to which each   link of the lattice belongs with probability $p$.
Now we define the $p$-clusters as the connected subsets of $\mathcal{O}\cup \mathcal{O}'$.
The original definition corresponds to $p=0$.
As in standard percolation problems,  it is evident that, regardless of where we are in the phase diagram, the $p$-clusters will percolate if $p$ is made sufficiently large  \cite{aharony2003introduction}. 
Therefore, if $p$ is made sufficiently large, the percolation transition line will preempt the Higgs line, giving a part of the deconfined phase in which $\<Q\>\neq 0$.

In principle it is possible to modify the lattice action to take account of the above modification to the definition of clusters. 
We do not go into this here, but a heuristic picture is that increasing $p$ would correspond to
adding terms depending only on $Q$, in particular to 
decreasing the bare mass-squared of the field $Q$.\footnote{Since the field $Q$ vanishes in the replica limit $N\to 1$, changing the bare mass of $Q$ (or adding other terms to the action that depend only on $Q$) 
has no effect on conventional  thermodynamic observables. However it affects replica correlators such as (\ref{eq:pconn}).}

The original topology of the phase diagram, in which the percolation transition lies on top of the Higgs line, is \textit{stable} for small enough $p$: roughly speaking,
this topology arises if the bare mass of $Q$ is sufficiently large.
However if $p$ is increased enough, 
the bare mass of $Q$ may become (sufficiently) negative, in part of the phase diagram, to enter the pair condensate phase.

Leaving aside the heuristic replica discussion, the stability of the phase diagram topology in Fig.~\ref{Fig:phasediagramwithpercolation} (with the percolation line on top of the Higgs line) 
relies formally on  the geometrical ``two-cluster'' operator \cite{vasseur2012logarithmic} being renormalization-group irrelevant at the Higgs transition and at the multicritical point \cite{somoza2021self}.
The irrelevance of this operator means that the critical clusters 
look  at large scales like topologically one-dimensional \textit{loops} 
which rarely revisit the same location (self-contacts become rare after coarse-graining of the clusters). 
This absence of self-contacts, at large scales, means that sufficiently small changes to way the clusters are defined (such as a small increase of $p$ above zero, in the modification described above) do not change the large scale geometry of the clusters, and do not move the percolation transition line away from the Higgs line.

The irrelevance of the two-cluster operator can be inferred from the fact that the percolation and Higgs transitions are found numerically to coincide. 
The emergent loop-like structure is  also visually apparent in snapshots of configurations \cite{somoza2021self}.
(The fact that the fractal dimension of the critical clusters is small (around $d_f=1.73$ at the Higgs transition, and around $d_f=1.77$ at the multicritical point \cite{somoza2021self}) 
also suggests that branching structures are likely to be rare.) 
But for a more direct demonstration, the irrelevance of the two-cluster operator could also be tested numerically by computing an appropriate geometrical correlation function \cite{vasseur2012logarithmic}.

In this Appendix we have discussed two ways for holes to percolate: $(i)$ via a thermodynamic transition, i.e. along the Higgs line or at the MCP where the Higgs line terminates; and $(ii)$ via a purely geometrical transition with no thermodynamic significance.\footnote{Note that the latter kind of transition occurs in the small-$x$ part of the phase transition, see Fig.~\ref{Fig:phasediagramwithpercolation}.}
We note finally that the fractal properties of the critical clusters are very different at the two kinds of transition. In case $(i)$ the critical clusters have a fractal dimension smaller than two, and resemble non-self-intersecting loops on large scales. In case $(ii)$ they resemble conventional percolation clusters \cite{aharony2003introduction}, with a fractal dimension larger than 2.5. 
This is one way to determine whether (in a more general model) the patching scheme is breaking down simply because of a poor definition of clusters, or for the more fundamental reason that the Higgs line is being crossed.

\section{Field theory approaches to the multicritical point}
\label{app:fieldtheoryissues}

This appendix is mostly concerned with the multicritical point  of $\mathbb{Z}_2$ gauge theory with $\mathbb{Z}_2$ matter, and possible field theory approaches to it. 
We  spell out in more detail brief comments made in \cite{somoza2021self}  about continuum approaches to the MCP based on Chern-Simons theory.  In addition, we briefly list some more speculative field theory approaches to this critical point and related ones, in particular an approach based on analytic continuation in the rank of the global symmetry group.

\subsection{Mutual Chern-Simons Theory}

 Let's first review the challenges to  constructing a continuum field theory description of the self-dual multicritical point. 
 (We distinguish continuum field theories from lattice field theories such as Eq.~\ref{eq:Z3Dgaugerep}.)
 
The reasons why the MCP cannot be described by Landau theory or by a ``Landau$^*$'' theory have been discussed in the main text and in \cite{somoza2021self}.
Another possible approach is via mutual Chern-Simons theory 
\cite{hansson2004superconductors,diamantini2008topological,kou2008mutual,mcgreevy2017tasi}. The challenge in this description  arises from the need to include monopole operators in the action. This was mentioned  briefly in  Ref.~\cite{somoza2021self} and we give a little more detail here.

First consider a self-dual  mutual Chern-Simons action, involving a pair of $U(1)$ gauge fields $a^{1,2}$ and a pair of complex scalar fields, $z^{1,2}$ \cite{hansson2004superconductors,diamantini2008topological,kou2008mutual,mcgreevy2017tasi}:
\ba\notag
\mathcal{L}_{\rm CS}^{(k)} = & \,
 i \f{k}{2\pi} \epsilon_{\mu\nu \lambda} a^1_\mu \partial_\nu a^2_\lambda +  \f{1}{2} \sum_{I = 1,2}  \hspace{-0.5mm}
 |(\partial - i a^I) z^I|^2  
 \\ + & \f{m^2}{2}  \hspace{-0.8mm} \sum_{I = 1,2}
 \hspace{-0.5mm} | z^I|^2
  +  
 \lambda \hspace{-0.5mm} \lf |z^1|^4 + |z^2|^4 \ri 
 + \lambda'   |z^1|^2 |z^2|^2
 + \ldots
 \label{eq:mutualCS}
 \end{align}
 (in Euclidean signature). The ellipses  stand for other local couplings that we could add.
 
When the matter fields are massive, this theory describes a certain state with $\mathbb{Z}_k$ topological order \cite{hansson2004superconductors,diamantini2008topological,kou2008mutual,mcgreevy2017tasi}
(so that the case of main interest here is $k=2$).
The elementary ``e'' and ``m'' anyons are associated with quanta of $z^1$ and $z^2$ respectively. 
By tuning the renormalized mass-squared $m^2$ to zero, it is possible to access a phase transition (perhaps first order \cite{geraedts2012monte}) between this deconfined phase and a trivial phase. 

However, this  phase transition is not the one that we want. 
The reason is that the above field theory (for $k=2$) does not have the same symmetry as the problem of interest.
Eq.~\ref{eq:mutualCS} 
has, among other symmetries, a $U(1)\times U(1)$ \textit{global} symmetry  that arises from conservation of gauge flux for each of the two gauge fields.
In other words, the deconfined phase described by the above Lagrangian is a different symmetry-enriched topological order to that in the $\mathbb{Z}_2$ gauge theory with $\mathbb{Z}_2$ matter.
The e and m anyons are charged under the first and second U(1) groups, respectively. 
Note that, in this theory,  the e and m anyons have  distinct antiparticles. For example, e has an antiparticle e$^*$:  e and e$^*$ are topologically equivalent in the case $k=2$,
but carry opposite U(1) charge  of $+1/2$ and $-1/2$ units respectively. (Analogously in the $m$ sector.)

In order to cure this problem, we must allow for the terms in the action which reduce the symmetry to the  physical symmetry: i.e. terms which break the two $U(1)$ symmetries. 
Physically, this corresponds to allowing processes in which, for example, an e anyon transforms into its antiparticle e$^*$ (so that the distinction between particle and antiparticle disappears at large scales).
Since the symmetries in question correspond to flux conservation of the two gauge fields, these processes are Dirac monopoles (instantons):  
pointlike sources of gauge flux in spacetime.
The necessity of allowing such instanton events, in order to remove conserved currents that are unphysical in the present context
was pointed out by  Hansson et al. in Ref.~\cite{hansson2004superconductors}.

Allowing monopoles in the path integral is equivalent to adding monopole operators  to the action   (for discussions of monopole operators in gauge theories see e.g. 
\cite{borokhov2003monopole,metlitski2008monopoles} and references therein)
\ba\notag
\mathcal{L} = & \,
 i \f{2}{2\pi} \epsilon_{\mu\nu \lambda} a^1_\mu \partial_\nu a^2_\lambda +  \f{1}{2} \sum_{I = 1,2}
 |(\partial - i a^I) z^I|^2  
 \\
 &
 + \f{m^2}{2}   \sum_{I = 1,2} | z^I|^2
  +  
 \lambda \lf |z^1|^4 + |z^2|^4 \ri 
 + \lambda'   |z^1|^2 |z^2|^2
 \notag  \\
 & + \kappa \sum_{I=1,2} 
\lf  \mathcal{M}^{(I)} +  \mathcal{M}^{(I)*}   \ri.
 \label{eq:mutualCS2}
 \end{align}
Here $\mathcal{M}^{(I)}$ (respectively $\mathcal{M}^{(I)*}$)
inserts a minimal-strength Dirac monopole (antimonopole).

It is possible that this UV theory
(for appropriate small values of the couplings, and when the renormalized mass is tuned to zero)
flows to the same fixed point as the MCP in the $\mathbb{Z}_2$ gauge theory with $\mathbb{Z}_2$ matter, 
i.e. in the Wegner/Fradkin-Shenker model.
Moving along the self-dual line would correspond, heuristically, to changing $m^2$, and  perturbing away from the self-dual line would correspond to making the masses different for $z^1$ and $z^2$, i.e. to adding a term ${|z^1|^2-|z^2|^2}$.

Unfortunately, however, at present it is challenging to deal with field theories such as Eq.~\ref{eq:mutualCS2}, so Eq.~\ref{eq:mutualCS2} is not --- yet --- a predictive theory for the MCP.
While $\mathcal{M}^{(I)}$ and $\mathcal{M}^{(I)*}$ are well-defined operators in the continuum theory, 
they are not local expressions in terms of the fields $z^I$ and $a^I$,
so this action is far from the domain where simple mean-field like reasoning is a useful starting point.
For example, simply by looking at Eq.~\ref{eq:mutualCS2}, 
it is not easy to say whether there should be a second order transition, or what the topology of the phase diagram in the vicinity of this transition should be.

However, the above field theories and extensions of them can certainly be useful for other critical points involving gapless anyons, for example if the monopole operators become irrelevant, which is expected to happen at large enough $k$ \cite{chongunpublished},
or if the U(1)$\times$U(1) symmetry exists from the outset as a physical symmetry of the problem. 
We comment further on the case of large $k$ below, since this can give models where braiding effects on mutually nonlocal gapless anyons can be understood relatively simply.

\subsection{Loops and braiding phases}

In order to motivate some other approaches below, let us review a geometrical way of rewriting the partition function for the $\mathbb{Z}_2$ gauge theory on the self-dual line. This is as a loop gas in three Euclidean dimensions.
Here we will be schematic
 --- for more details see the discussion in Ref.~\cite{somoza2021self} and the related loop gas in \cite{geraedts2012monte}. 

The partition function of the loop gas involves a sum over configurations two kinds of unoriented loops, one representing e worldlines and one representing m worldlines. 
The e worldlines live on a cubic lattice $\mathfrak{L}$, and the m worldlines live on the dual\footnote{In the loop representation of Eq.~\ref{eq:Z3Dgaugerep}, these loops may have self-intersections, but we do not expect this to be essential (i.e.  we believe that these self-intersections are irrelevant at large scales --- see the discussion at the end of  App.~\ref{sec:percolationanalytic}). It is also possible to formulate models where both kinds of loops live on the same lattice, see Sec.~X of Ref.~\cite{somoza2021self}.} cubic lattice $\mathfrak{L}'$ (whose sites are at the centers of cubes of $\mathfrak{L}$). 
The Boltzmann weight consists of (i) local interactions between worldlines, and (ii) a topological contribution of the schematic form $(-1)^{X}$, where ${X=X(\mathcal{C}_{\rm e}, \mathcal{C}_{\rm m})}$ is the $\mathbb{Z}_2$ linking between e worldlines and m worldlines.  
Schematically,
\be\label{eq:unorientedloopsum1}
Z_{\rm IGT} = \sum_{\mathcal{C}_{\rm e}, \mathcal{C}_{\rm m}} W(\mathcal{C}_{\rm e}, \mathcal{C}_{\rm m}) (-1)^{X(\mathcal{C}_{\rm e}, \mathcal{C}_{\rm m})},
\ee
where e.g. $\mathcal{C}_e$ denotes the configuration of unoriented $e$ loops and $W(\mathcal{C}_{\rm e}, \mathcal{C}_{\rm m})$ contains the local interactions between loop segments.\footnote{For more careful discussion of boundary conditions, etc. see \cite{somoza2021self}.}

Note that the topological term is the difference from a Landau theory.
If we dropped this term, the resulting partition function 
\be\label{eq:unorientedloopsumLGW}
Z_{\rm L} = \sum_{\mathcal{C}_{\rm e}, \mathcal{C}_{\rm m}} \widetilde{W}(\mathcal{C}_{\rm e}, \mathcal{C}_{\rm m}),
\ee
potentially with addditional local interactions between the loop segments,
is a ``worldline'' representation of a more conventional lattice theory for two real Landau-like order parameters $\phi_x$ and $\phi_y$, one living on each sublattice.
Depending on the microscopic interactions, this theory can have a transition in the XY universality class, with an emergent O(2) symmetry for the order parameters (see the discussion in \cite{somoza2021self}).

First let us note why, in this language, we expect the exponents of the MCP to be different from XY exponents.
We may think about Eq.~\ref{eq:unorientedloopsum1} in terms of a heuristic real space renormalization group, in which we coarse-grain the loop configurations. We expect that the topological linking term in the Boltzmann weight is preserved during coarse-graining, but  the local interactions in $W$ will be renormalized. As a result of the 
linking phase, the RG transformation for $W$ in 
Eq.~\ref{eq:unorientedloopsum1}
will be different from that for $\widetilde W$ in 
Eq.~\ref{eq:unorientedloopsumLGW}.
As a result, the fixed point and its exponents, for the case of the gauge theory MCP, should be expected to be different from those of the XY model (even if they are numerically close).

Nevertheless, the fact that the exponents are close suggests that the effect of the topological phase on the RG transformation for $W$, in the vicinity of the fixed point, may be quantitatively mild in some sense. 
However, in the $\mathbb{Z}_2$ case it is not obvious how this should be quantified (unless it was possible to construct the real-space RG transformation numerically).

One regime where the effect of the braiding phases could be quantified is for the $\mathbb{Z}_k$ problem at large $k$.
In this case, the natural loop model involves \textit{oriented} loops
(worldlines of the elementary anyons), and the braiding phase is $e^{2\pi i Y/k}$, where $Y$ is now the signed linking number that takes account of the orientation of the loops.
Since this phase becomes small at large $k$, the RG transformation law will converge to that of a simple Landau theory (now for two complex order parameters) at large $k$.\footnote{We have discussed with Chong Wang one formal way to make this precise. The linking phase may be expressed as  a double integral over the currents $J^{(1)}$ and $J^{(2)}$ of the two complex fields. 
(Modulo subtleties about the choice of regularization, this is equivalent to formally integrating the gauge fields out of the Chern-Simons action in Eq.~\ref{eq:mutualCS}.)
Then formally we could do perturbation theory in $1/k$, corresponding to an expansion in correlation functions of current operators in the initial Landau theory.
As a toy model, we could consider a fine-tuned critical point, in which the ``parent'' Landau theory fixed point was Gaussian.}

The Chern-Simons theory at large $k$ has  been treated much more completely than the above schematic discussion in a new preprint by Shi and Chatterjee \cite{shi2024analytic} that appeared after this work was completed. Ref.~\cite{shi2024analytic}  uses a ``large $N$'' approach to the mutual Chern-Simons theory to compute exponents as a function of $k/N$. 
(Large $N$ has also been considered in unpublished work by Ye and Wang \cite{chongunpublished}.)
It was  noted there that the exponents converge to those of Landau theory at large $k$, and also that this fact does not rely on large~$N$.

\subsection{A replica field theory approach to the MCP?}

The loop gas picture also suggests the following amusing but very speculative possibility. 
Above, starting with the Lagrangian in Eq.~\ref{eq:mutualCS},
we were forced to add monopole operators in order to remove 
an unwanted ``multiplicity'' of the anyons: 
the theory in Eq.~\ref{eq:mutualCS} has two degenerate quasiparticles in the e sector, namely e and e$^*$, 
whereas at low energies we wanted only a single anyon in this sector  (and similarly for the m sector).
Here we consider the possibility of removing the unwanted multiplicity \textit{without} using monopoles, using the replica trick instead.
(This should not be confused with a very different application of the replica trick in App.~\ref{sec:percolationanalytic}.)

We will motivate this possibility using a slightly unconventional lattice regularization of the field theory in (\ref{eq:mutualCS}), in which it becomes a simple loop gas (similar to that in  Ref.~\cite{geraedts2012monte}).\footnote{Essentially we regularize the matter fields on a lattice (more precisely, we place $z^1$ on a cubic lattice,  and $z^2$ on the dual cubic lattice) but we keep the gauge fields in the continuum. Then integrating the gauge fields out gives the purely topological interaction between the worldlines.}
In this loop gas we again have two kinds of worldlines, representing e and m loops, with a linking phase, but now each loop carries an orientation which distinguishes particle worldlines from antiparticle worldlines. We choose a lattice regularization  in which the local interactions between the loops are independent of the loop orientations.

With this choice of lattice regularization, it is possible to perform the sum over loop orientations explicitly in the partition function, to give a topological contribution  $2^{\text{no. loops}}$ to the Boltzmann weight:
\be\label{eq:unorientedloopsum2}
Z^{(2)} = \sum_{\mathcal{C}_e, \mathcal{C}_m} W(\mathcal{C}_e, \mathcal{C}_m) (-1)^{X(\mathcal{C}_e, \mathcal{C}_m)}2^{\text{no. loops}},
\ee
The sum is now over unoriented loops, as in (\ref{eq:unorientedloopsum1}).
The factor of $2^{\text{no. loops}}$ is the remaining difference between this partition function and the one we wish to study for the original MCP in Fig.~\ref{Fig:phasediagram}.

The partition function above naturally generalizes to
\be\label{eq:unorientedloopsumn}
Z^{(n)} = \sum_{\mathcal{C}_e, \mathcal{C}_m} W(\mathcal{C}_e, \mathcal{C}_m) (-1)^{X(\mathcal{C}_e, \mathcal{C}_m)} (2n)^{\text{no. loops}}.
\ee
for an arbitrary integer $n\geq 1$.
In turn, this loop gas  may be obtained from a lattice analog
 of a mutual Chern-Simons theory like that in Eq.~\ref{eq:mutualCSn}, but for scalar fields ${{\mathbf z}^I=(z^I_1,\ldots, z^I_n)}$ with $n$ flavors:
\ba
\mathcal{L}_{\rm CS}^{(2,n)} = & \,
 i \f{2}{2\pi} \epsilon_{\mu\nu \lambda} a^1_\mu \partial_\nu a^2_\lambda +  \f{1}{2} \sum_{I = 1,2}
 |(\partial - i a^I) {\mathbf z}^I |^2  
+ \ldots
 \label{eq:mutualCSn}
 \end{align}
 
We see that partition function in Eq.~\ref{eq:unorientedloopsumn} becomes equal to the desired partition function of $\mathbb{Z}_2$ gauge theory with $\mathbb{Z}_2$ matter if we formally take the replica limit
\be
n\to 1/2.
\ee
Therefore it is tempting to ask whether there is some way to access the desired multicritical point
using an ``${n\to 1/2}$'' replica limit of the multiflavor continuum theory 
in Eq.~\ref{eq:mutualCSn}.
 
At this point, it is important to note that  it is not a priori obvious whether these lattice theories have the same infrared behavior as the continuum field theories (when the latter are regularized in a more conventional way). 
In fact, as discussed below, this is connected to a question of  a nontrivial emergent  symmetry.

A more generic lattice version of the continuum Chern-Simons theory 
will not have such a simple form as (\ref{eq:unorientedloopsumn}). 
The key feature that was used in obtaining Eq.~\ref{eq:unorientedloopsumn} was that the local interactions between loop  segments were independent of their relative orientations. As a result, we were able simply to sum over the orientations to produce the factor of $2^\text{no. loops}$.

This is a hint that Eq.~\ref{eq:unorientedloopsumn} encodes an enlarged symmetry compared to a generic lattice regularization of the Chern-Simons theory. 
We can see this by noting that the loop gas (\ref{eq:unorientedloopsumn})
can also be obtained 
from a lattice $\mathbb{Z}_2$ gauge theory 
with an explicit 
${\mathrm{so}(2n) \times \mathrm{so}(2n)}$ 
continuous global symmetry.\footnote{\label{mutualZ2footnote} One  representation of the standard $\mathbb{Z}_2$ gauge theory with $\mathbb{Z}_2$ matter is in terms of a  lattice ``BF'' action \cite{hansson2004superconductors,senthil2000z}. This takes the schematic form 
\be\label{eq:mutualZ2}
\mathcal{S} = 
- J \sum_{\<ij\>\in \mathfrak{L}} \sigma_{ij} \tau_i \tau_j
- J' \sum_{\<ij\>\in \mathfrak{L}'} \widetilde{\sigma}_{ij} \widetilde{\tau}_i \widetilde{\tau}_j + \mathcal{S}_\text{top}(\sigma, \widetilde\sigma),
\ee
where $\sigma$ is a $\mathbb{Z}_2$ gauge field on the cubic lattice $\mathfrak{L}$ and $\widetilde{\sigma}$ is the dual gauge field, living on the dual cubic lattice $\mathfrak{L}'$ (similarly for the matter fields $\tau$ and $\widetilde{\tau}$), and $\mathcal{S}_\text{top}(\sigma, \widetilde\sigma)$ is a topological action \cite{senthil2000z,hansson2004superconductors}.
Integrating out $\widetilde{\sigma}$ and $\widetilde{\tau}$ shows that this formulation is equivalent to the standard $\mathbb{Z}_2$ gauge theory plus $\mathbb{Z}_2$ matter (Eq.~\ref{eq:Z3Dgaugerep}), with $e^{-2K} = \tanh J'$.
Now, we can extend the theory to one with $2n$ flavours simply by replacing $\tau$ with a unit-length vector $\vec{\tau} = (\tau_1,\ldots, \tau_{2n})$ and similarly for $\widetilde{\vec{\tau}}= (\widetilde{\tau}_1,\ldots, \widetilde{\tau}_{2n})$:
 \be\label{eq:Z2topoactionOn}
 \mathcal{S} = 
- J \sum_{\<ij\>\in \mathfrak{L}}
 \sigma_{ij} \vec{\tau}_i . \vec{\tau}_j
- J' \sum_{\<ij\>\in \mathfrak{L}'} 
\widetilde{\sigma}_{ij} 
\widetilde{\vec{\tau}}_i .\widetilde{\vec{\tau}}_j + 
\mathcal{S}_\text{top}(\sigma, \widetilde\sigma). 
 \ee 
 For a slightly simpler strong-coupling expansion it is convenient to modify the  Boltzmann weight to the Nienhuis-like \cite{nienhuis2010loop} form
 \be\label{eq:Z2topoactionnOn2}
e^{-\mathcal{S}} = 
e^{\mathcal{S}_\text{top}(\sigma, \widetilde\sigma)}
\prod_{\<ij\>\in \mathfrak{L}}(1+y \sigma_{ij} \vec{\tau}_i . \vec{\tau}_j) 
\prod_{\<ij\>\in \mathfrak{L}}(1+y' \widetilde{\sigma}_{ij} 
\widetilde{\vec{\tau}}_i .\widetilde{\vec{\tau}}_j) 
 \ee
 The strong coupling expansion of this theory maps it to  a loop gas similar to that in Eq.~\ref{eq:unorientedloopsumn}, except that loops can visit a given node more than once (a given link is visited at most once). 
 (This can be seen from standard considerations for loop models \cite{cardy1996scaling} together with  basic properties of $\mathcal{S}_{\text{top}}$ \cite{hansson2004superconductors}.)
 Note that both (\ref{eq:Z2topoactionOn}) and (\ref{eq:Z2topoactionnOn2}) have manifest ${[O(2n)/\mathbb{Z}_2]\times [O(2n)/\mathbb{Z}_2]}$ global symmetry. Here the two $O(2n)$ groups represent rotations/reflections of $\vec{\tau}$ and $\widetilde{\vec{\tau}}$ respectively (and we have quotiented out the gauge symmetries associated with $\sigma$ and $\widetilde{\sigma}$ respectively).}
This is larger than the explicit continuous global symmetry of the 
continuum Chern-Simons Lagrangian, which is ${[\mathrm{su}(n)\times u(1)] \times [\mathrm{su}(n)\times u(1)]}$.
(The $\mathrm{su}(n)$ factors are flavor symmetries, for the e and m sectors respectively, and the $u(1)$ factors are the flux conservation symmetries discussed above. For simplicity we consider only the continuous part of the symmetries, neglecting discrete factors.)

In order for the replica limit of the continuum theory to make sense, what we would probably require is the following. 
Starting with Eq.~\ref{eq:mutualCSn} with appropriate quartic couplings, 
we would hope to find --- somewhere in the parameter space\footnote{It would not necessarily matter if such  a fixed point  was highly unstable (or even complex \cite{gorbenko2018walking})   for $n>1/2$, since   all that would matter would be the number of (symmetry-allowed and nonvanishing) relevant perturbations when $n=1/2$, and the unitarity of the theory (in the IR) for this value of $n$.} --- a nontrivial fixed point where ${\rm so}(2n)\times {\rm so}(2n)$ global symmetry \textit{emerged} in the infrared.
Then, we would hope to analytically continue the critical exponents of this fixed point to $n=1/2$.

This scenario passes a basic consistency check. 
The number of conserved current operators for a theory with ${\rm so}(2n)\times {\rm so}(2n)$ is $2n(2n-1)$,\footnote{This the number of Lie group generators.} which of course tends to zero when $n\to 1/2$: this is as as desired, since (a priori) there is no reason to expect the MCP to have any continuous symmetry.

By contrast, it is unlikely  that a fixed point of $\mathcal{L}^{(2,n)}$ \textit{without} any symmetry enhancement could be sensibly continued to ${n=1/2}$.
A theory with only 
${[\mathrm{su}(n)\times u(1)] \times [\mathrm{su}(n)\times u(1)]}$ global symmetry has ${2(n^2-1) + 2}$ conserved current operators. 
When ${n\to 1/2}$, this number tends to $1/2$, which is unlikely to have a sensible interpretation.

Note that if ${\rm so}(2n)\times {\rm so}(2n)$ symmetry did emerge, 
for some choice of couplings in the Chern-Simons theory, 
it would not act locally on the fields $z^I$ and $a^I$.
For example, consider the $n^2-1$ operators 
${\mathcal{O}_{ab}^{(1)} = z^1_a z^{1*}_b- \f{1}{n} |\vec{z}^1|^2 \delta_{ab}}$, 
which transform in the adjoint representation of ${\rm su}(n)$,
together with the $n(n+1)/2$ monopole operators
$\mathcal{M}^{(2)}_{ab}$,
which transform in the symmetric tensor representation of $\mathrm{su}(n)$, 
together with their complex conjugates 
$\mathcal{M}^{(2)*}_{ab}$.
All of these correspond to operators which transform an e particle of one species into an e particle of another species, 
so they would presumably unite to form a single (symmetric tensor) representation of ${\rm so}(2n)$, of dimension $2(n+1)(n-1/2)$.\footnote{The corresponding operators in the theory (\ref{eq:Z2topoactionOn}) of footnote~\ref{mutualZ2footnote} would be of the form $\tau^1_{a}\tau^1_b - \f{1}{2n} \delta_{ab} (\vec{\tau}^1)^2$.}

The above scenario for the RG flows is speculative. If it did hold, the advantage would be that we have eliminated the need to include monopoles in the continuum action: the unwanted symmetry is eliminated using the replica limit, rather than using monopoles.
However, we would still need an analytic method
 of studying the hypothetical fixed point in which $n$ could be treated as a continuously variable parameter\footnote{The large $n$ expansion is unlikely to be useful: monopole operators have large scaling dimensions at the fixed points that are easily accessible at large $n$ \cite{murthy1990action}, so the  enhancement of symmetry between the $\mathcal{O}_{ab}$ and $\mathcal{M}_{ab}$ operators described in the previous paragraph is unlikely at such fixed points.} (for example, fixed-dimension RG).
So it is not yet clear whether this approach could be useful. 
Nevertheless, this kind of (putative) symmetry enhancement in Chern-Simons theory is interesting in its own right even for fixed $n$.

\bibliography{IGTrefs.bib}
\end{document}